\documentclass[twocolumn,showpacs,aps,amssymb,floatfix,prd,amsmath,preprintnumbers]{revtex4} 
\usepackage{graphicx}
\usepackage{dcolumn}
\usepackage{bm}
\begin{document}
\def \reals{{\mathbb R}}
\def\be{\begin{equation}}
\def\ee{\end{equation}}
\def\bea{\begin{eqnarray}}
\def\eea{\end{eqnarray}}
\def\nn{\nonumber}
\def\th{\theta}
\def\ph{\phi}
\def\lt{\left}
\def\rt{\right}
\def\degree{\mathop{\rm {{}^\circ}}}
\input epsf.tex
\title{Time delay and magnification centroid due to   gravitational   lensing by black holes and naked singularities}

\author{K.~S.~Virbhadra}
    \email[Email address : ]{shwetket@yahoo.com}
    \affiliation{Department of Mathematics, Physics and Statistics, University of the Sciences in Philadelphia, Philadelphia, PA 19104}

\author {C.~R.~Keeton}
   \email[Email address : ]{keeton@physics.rutgers.edu}
    \affiliation{Department of Physics and Astronomy, Rutgers University, 136 Frelinghuysen Road, Piscataway, NJ 08854 }
\begin{abstract}
We model the massive dark object at the center of the Galaxy as a Schwarzschild black hole as well as 
Janis-Newman-Winicour naked
singularities, characterized by the mass and  scalar charge  parameters, and study  gravitational lensing 
(particularly time 
delay, magnification centroid, and total  magnification)  by them. We find that the 
lensing features are qualitatively similar (though  quantitatively different) for the Schwarzschild black 
holes, weakly naked, and 
marginally strongly naked singularities. However, the lensing  characteristics of strongly naked 
singularities are qualitatively very different from those due the Schwarzschild black holes.
The images produced by  Schwarzschild black hole lenses and weakly naked and marginally strongly naked singularity lenses
always have positive time  delays.
On the other hand, the strongly naked singularity lenses can give rise to images with positive, zero, or negative 
time delays. 
In particular, for a large angular source position  the direct image (the outermost image on the same side 
as the source) due to strongly naked singularity lensing  always  has negative time delay.
We also found  that the  scalar
field  decreases the time delay and increases the magnitude of magnifications of images; 
this result could have important implications for cosmology.
As the Janis-Newman-Winicour metric also describes the exterior gravitational field of a scalar star, naked 
singularities as well as scalar star lenses, if these exist
in nature, will serve as more efficient cosmic telescopes than  regular gravitational lenses. 
\end{abstract}

\pacs{95.30.sf, 04.20.Dw, 04.70.Bw, 98.62.Sb }

\maketitle

\section{\label{sec:Intro}Introduction}

A naked (visible) singularity is defined as a spacetime singularity which can be seen by some observer and
also lies to the future of some point of the spacetime \cite{Pen98}. The well-known weak cosmic 
censorship hypothesis (WCCH) of Penrose essentially states  that, generically, spacetime  
singularities of physically realistic gravitational collapse are hidden within black holes \cite{Pen98,Wal97}.
The concept of visible singularities is objectionable to many scientists, as their existence
is thought to have alarming  astrophysical implications. On the other hand, a failure of the WCCH will 
give us the great opportunity  to probe the extremely strong gravitational fields that will help in the 
discovery of the physical laws of quantum gravity. Despite many industrious efforts,  we are still far from 
having a general proof (or disproof) of this 
hypothesis,  and Penrose \cite{Pen98} expected that  radically new mathematical techniques might be required 
for this purpose. As a proof or  disproof  of this hypothesis appears to be  inordinately difficult, 
it may be easier to find a persuading  counterexample to demonstrate that the hypothesis 
is not correct.  Numerous diligent efforts have been put in this direction in the last four decades; however,  
we still do not have a single convincing counterexample  to the  WCCH (see references in 
 \cite{Pen98,Wal97,Vir99}). In a seminal review,  Penrose \cite{Pen98} concluded that  the question of the 
cosmic censorship  is still very much open and considered  this to be possibly the  most important
unsolved problem in classical general relativity.

Given that we  have neither a proof (or disproof) nor a convincing counterexample  of the 
WCCH, it is 
important to explore whether or not this hypothesis could be tested observationally. 
To this end, Virbhadra {\it et al.} \cite{VNC98} introduced  a theoretical research project using the gravitational
 lensing phenomena and
encouraging results came out of that. Further, Virbhadra and Ellis \cite{VE00} obtained
 a new gravitational lens
equation that allows  large deflection of light and therefore it can be used 
to study strong gravitational
field lensing. They used this lens equation to study the gravitational lensing due to 
light deflection close to the
photon sphere of the supermassive ``black hole'' at the center of the Galaxy. They found 
that the presence of a photon 
sphere gives rise to   a theoretically infinite sequence of  highly demagnified images
 on both sides of the optical 
axis (the line joining the lens and the observer) and they termed these {\em relativistic 
images}. Virbhadra and Ellis
\cite{VE02} further extended the previous studies of Virbhadra {\it et al.} \cite{VNC98} 
in detail. They also organized
the investigations by classifying   naked singularities in two groups: {\em weakly naked} 
and {\em strongly naked}
singularities. They modeled  massive dark objects at the  centers of a few galaxies 
as  Schwarzschild black 
holes and Janis-Newman-Winicour weakly as well as strongly naked singularities, and 
studied  gravitational lensing by  them. The Schwarzschild black holes as well as weakly naked
singularities have qualitatively similar lensing characteristics : Both have one Einstein
 ring and no radial critical curves.
On the other hand, the strongly naked singularities have  qualitatively different lensing 
features, i.e., they give rise to
two  or none  Einstein rings and one radial critical
curve.  After publication of these results 
\cite{VNC98, VE00,CVE01,VE02}, there has been  a growing curiosity in black hole lensing and many 
interesting papers have appeared in  last few years (see \cite{Per,Zaketal,Amoreetal, Bozzaetal,
 Eioraetal, KNPF, MajuMukh,KP05, KPIW, ExoticLens, BHLens, Scalar} and references 
therein).

In this paper, we study the time delay, magnification centroid, and total magnification of images
due to  gravitational lensing by Schwarzschild black holes and  Janis-Newman-Winicour naked 
singularities. One of the most striking results in this paper is that the strongly naked 
singularities can give rise to images  with negative time delays.  We use geometrized units (i.e.,
$G=1, c=1$) throughout this paper; however, we finally compute  time delays in  terms of minutes.
We use MATHEMATICA \cite{Math} for computations.


\section{\label{sec:LE&DE} Lens Equation, Light Deflection Angle, and Classification of Naked Singularities}

In this Section, we write  in brief some of the results obtained in previous papers
(\cite{VNC98,VE00,CVE01,VE02,Vir97,VJJ97})  and refine the  classification of
naked singularities given in \cite{VE02}, because these are
required for computations and analysis of results in this paper. 

Virbhadra and Ellis \cite{VE00} derived a gravitational lens equation  that  permits 
small as well as large  bending angle of light,  and  that is given by

\begin{equation}
  \tan\beta =  \tan\theta - \alpha ,
  \label{GravLensEqn}
\end{equation}
with
\begin{equation}
  \alpha \equiv
    \frac{D_{ds}}{D_s}  \lt[\tan\theta + \tan\lt(\hat{\alpha} - \theta\rt)\rt] \text{.}
   \label{Alpha}
\end{equation}
$D_{s}$, $D_{ds}$ and $D_{d}$, respectively, are  the observer-source, the lens-source,  
and  the observer-lens distances.  $\hat{\alpha}$ is the light  bending angle.  $\theta$ and
$\beta$ are, respectively,   angular positions of an image and  an unlensed source  measured 
from the optical  axis. (See Fig. 1 in \cite{VE00}.) The impact  parameter $ J = D_d \ \sin\theta$.
For small angles, the   Eq. (1) reduces to the most well-known lens equation  used for studying 
lensing in a weak gravitational field \cite{SFE92}.

In  circularly symmetric gravitational lensing, the  magnification $\mu$ of an 
image is
\begin{equation}
  \mu = \lt( \frac{\sin{\beta}}{\sin{\theta}} \ \frac{d\beta}{d\theta} \rt)^{-1} \text{.}
  \label{Mu}
\end{equation}
The tangential and radial magnifications are respectively expressed by
\begin{equation}
    \mu_t = \lt(\frac{\sin{\beta}}{\sin{\theta}}\rt)^{-1} ~ ~ ~ \text{and} ~ ~ ~ ~
    \mu_r = \lt(\frac{d\beta}{d\theta}\rt)^{-1}  \text{.}
    \label{MutMur}
\end{equation}
The singularities in $\mu_t$ and  $\mu_r$ in lens plane give, respectively, {\em tangential 
critical curves} (TCCs) and  {\em radial critical curves}  (RCCs), and their corresponding values
in the source plane are, respectively, known as the {\em  tangential caustic} (TC) and  {\em radial caustics}
(RCs).

Virbhadra {\it et al.}~\cite{VNC98} considered a general static and spherically symmetric spacetime  described the line element
\begin{equation}
ds^2 = B(r) dt^2 - A(r) dr^2 -  D(r) r^2 (d\vartheta^2 + \sin^2\vartheta d\varphi^2) 
    \label{GenMetric}
\end{equation}
and calculated the deflection angle $\hat{\alpha}\lt(r_0\rt)$  and impact parameter $J(r_0)$
for a light ray  with the  closest distance  of approach $r_0$. These are given by
\begin{equation}
    \hat{\alpha}\lt(r_0\rt)
              = 2 {\int_{r_0}}^{\infty}\lt(\frac{A(r)}{D(r)}\rt)^{1/2}
                       \lt[
                       \lt(\frac{r}{r_0}\rt)^2\frac{D(r)}{D(r_0)}\frac{B(r_0)}{B(r)}-1
                    \rt]^{-1/2} \frac{dr}{r}   - \pi
      \label{GenAlphaHat}
\end{equation}
and
\begin{equation}
    J\lt(r_0\rt)  = r_0  \sqrt{\frac{D(r_0)}{B(r_0)}} \text{.}
      \label{GenImpactPara}
\end{equation}
For $D\lt(r\rt) = 1$,  equations  $(  \ref{GenAlphaHat})$ and $(\ref{GenImpactPara}  )$  yield
the results obtained by Weinberg \cite{Wein72}.

The most general static and  spherically symmetric solution to the Einstein massless scalar
equations was independently obtained by Janis, Newman and Winicour as well as 
Wyman \cite{Janis-Newman-WinicourWyman}. As both solutions were available in  different coordinates, 
 they were  not known to be the same until Virbhadra \cite{Vir97} showed the
equivalence between the two by a  coordinate transformation. As Janis, Newman and Winicour
obtained this solution about thirteen years before Wyman, we prefer to call it the 
Janis-Newman-Winicour  solution.  Thus, the  Janis-Newman-Winicour  
solution (characterized by constant and real parameters, the ADM mass $M$ and the scalar 
charge $q$) is expressed by the line element
\begin{eqnarray}
   ds^2 &=&   \left(1-\frac{b}{r}\right)^{\nu} dt^2
      - \left(1-\frac{b}{r}\right)^{-\nu} dr^2 \nonumber \\
      &-& \left(1-\frac{b}{r}\right)^{1-\nu}  
      r^2 \left(d\vartheta^2  +\sin^2\vartheta \  d\varphi^2\right) 
    \label{Janis-Newman-WinicourMetric} 
\end{eqnarray}
and the  massless scalar field
\begin{equation}
    \Phi = \frac{q}{b\sqrt{4\pi}} \ln\left(1-\frac{b}{r}\right),
   \label{Phi}
\end{equation}
with 
\begin{equation}
   \nu = \frac{2M}{b} ~ ~ ~  \text{and } ~ ~ ~ b = 2 \sqrt{M^2+q^2}.
 \label{nub}
\end{equation}
This solution is  asymptotically Minkowskian and reduces to the Schwarzschild 
solution for $q=0$ (i.e., $\nu =1$). The Janis-Newman-Winicour  solution has a {\em globally naked strong curvature 
singularity} at $r = b$  for  all values of $q \neq  0$ and this solution is physically 
reasonable as  it satisfies the weak energy condition \cite{VJJ97}.
Virbhadra {\it et al.} obtained   the light deflection angle 
$\hat{\alpha}\left(r_0\right)$  for    large value of   $r_0$   (see equation (24) in \cite{VNC98}); we now 
re-express  that  using the equation $(\ref{nub})$, as follows:

\begin{eqnarray}
  \hat{\alpha}\left(r_0\right) = 2 \nu \left(\frac{b}{r_0}\right)
    &+& \left[ \nu(1-2\nu) + \pi\left(\nu^2-\frac{1}{16}\right) \right]
      \left(\frac{b}{r_0}\right)^2 \nonumber \\
     &+& \mathcal{O}{\left(\frac{b}{r_0}\right)}^3 \text{.}
  \label{VNCEq24nu} 
\end{eqnarray}


Virbhadra and Ellis \cite{VE00}  as well as Claudel {\it et al.}~\cite{CVE01} gave  two different definitions
of a photon sphere in a  static spherically symmetric spacetime. Both definitions gave  the same
results for a general static and spherically symmetric metric. Thus, according to both  definitions, 
the Janis-Newman-Winicour spacetime has only one  photon sphere  and that is situated at the radial distance \cite{VE00,CVE01}
\begin{equation}
   r_{ps} = \frac{b (1+2\nu) }{2}  \text{.}
   \label{Janis-Newman-WinicourPS}
\end{equation}
As $r=b$ is the curvature singularity, the photon sphere exists only for $\nu : 1/2 < \nu \leq 1$.

Defining 
\begin{equation}
  \rho = \frac{r}{b} , ~ ~ ~  \rho_0 = \frac{r_0}{b} 
  \label{rToRho}
\end{equation}
and using equations $(\ref{GenAlphaHat})$, $(\ref{GenImpactPara})$ and $(\ref{Janis-Newman-WinicourMetric})$, the 
deflection angle $\hat{\alpha}$ and the impact parameter $J$ for a light ray
in the Janis-Newman-Winicour spacetime are expressed in the form \cite{VNC98,VE00}

\begin{widetext}
\begin{equation}
   \hat{\alpha}\lt(\rho_0\rt) = 2 \  {\int_{\rho_0}}^{\infty}
   \frac{d\rho}{\rho \sqrt{1-\frac{1}{\rho}} \  \sqrt{\lt(\frac{\rho}{\rho_0}\rt)^2
   \lt(1-\frac{1}{\rho}\rt)^{1-2\nu} \lt(1-\frac{1}{\rho_0}\rt)^{2\nu-1}-1}} - \pi
    \label{Janis-Newman-WinicourAlphaHatRo}
\end{equation}
\end{widetext}

and
\begin{equation}
    J\lt(\rho_0\rt)  = 2M \frac{\rho_0}{\nu} \left(1-\frac{1}{\rho_0}\right)^{\frac{1-2\nu}{2} }
                         \text{\ .}
     \label{Janis-Newman-WinicourJRo}
\end{equation}
Obviously, equation $(\ref{Janis-Newman-WinicourPS})$ can now  be re-expressed as
\begin{equation}
   \rho_{ps} = \frac{ (1+2\nu) }{2}  \text{.}
   \label{Janis-Newman-WinicourPSRho}
\end{equation}

Equation  $(\ref{Janis-Newman-WinicourJRo})$ with  the expression for the impact parameter, $J=D_d \sin\theta$, give 
\begin{equation}
  \sin\theta   = \frac{2M}{D_d}
                 \frac{\rho_0}{\nu} \left(1-\frac{1}{\rho_0}\right)^{\frac{1-2\nu}{2} } \text{.}
\end{equation}
The  first derivative of the deflection angle $\hat{\alpha}$ with respect to $\theta$ is given by
\cite{VNC98,VE02}
\begin{equation}
\frac{d\hat{\alpha}}{d\theta} =  \hat{\alpha}'\lt(\rho_0\rt)
                               \frac{d\rho_0}{d\theta} ,
\label{DAlphaHatByDTheta}
\end{equation}
where
\begin{equation}
      \frac{d\rho_0}{d\theta} =
      \frac{\nu \rho_0 \left(1-\frac{1}{\rho_0}\right)^{\frac{1+2\nu}{2}}
             \sqrt{1-\frac{4}{\nu^2} \left(\frac{M}{D_d}\right)^2 {\rho_0}^2
             \left(1-\frac{1}{\rho_0}\right)^{1-2\nu} }    }
      {\frac{M}{D_d} \left(2\rho_0-2\nu-1\right)}
    \label{DRoByDTheta}
\end{equation}
and 
\begin{widetext}
\begin{equation}
    \hat{\alpha}'\lt(\rho_0\rt) = \frac{2\nu+1-2\rho_0}{{\rho_0}^2\lt(1-\frac{1}{\rho_0}\rt)}
                                  {\int_{\rho_0}}^{\infty}
                                  \frac{\lt(4 \nu \rho - 2 \nu - 1\rt) d\rho}
                                   {\lt(2\nu +1 - 2 \rho\rt)^2 \  \rho \ \sqrt{1-\frac{1}{\rho}} \  
                                    \sqrt{\lt(\frac{\rho}{\rho_0}\rt)^2
                      \lt(1-\frac{1}{\rho}\rt)^{1-2\nu} \lt(1-\frac{1}{\rho_0}\rt)^{2\nu-1} -1}} \text{.}
                  \label{Janis-Newman-WinicourAlphaHatPrime}
\end{equation} 
\end{widetext}
The prime denotes the  first derivative with respect to $\rho_0$.

Virbhadra and Ellis \cite{VE02} classified naked singularities in two groups:
{\em Weakly naked  singularities} (WNS) are those which are  contained within at least one  photon  sphere,
whereas {\em strongly naked  singularities} (SNS) are  those which are {\em not} covered within any  photon spheres. 
Therefore, according to this classification, the Janis-Newman-Winicour naked singularities  are strongly naked for
$0 \leq \nu \leq 1/2$ and weakly naked for  $1/2 < \nu < 1$.

For Schwarzschild black holes ($\nu=1$) as well as WNS ($1/2 < \nu < 1$), the deflection angle
 $\hat{\alpha}\lt(\rho_0\rt)$ monotonically increases with
the decrease in the closest distance of approach $\rho_0$ and $\hat{\alpha}\lt(\rho_0\rt) \rightarrow \infty$ as
$\rho_0 \rightarrow  \rho_{ps}$. As both have qualitatively similar  $\hat{\alpha}$ vs. $\rho_0$ graph,
their lensing features are also qualitatively similar \cite{VE00,VE02}. 
However, Virbhadra and Ellis \cite{VE02} missed noticing a point:    Though there are no photon 
spheres for  $\nu=1/2$,  the deflection angle behavior, according to the equation $(\ref{Janis-Newman-WinicourAlphaHatPrime})$,
is similar to the 
cases of the Schwarzschild  black holes and WNS;  therefore, their gravitational lensing features will be 
also qualitatively the same. In view of this, we now prefer to term $\nu = 1/2$ and
$0\leq\nu<1/2$ singularities, respectively, {\em marginally  strongly naked singularities} 
(MSNS) and  {\em strongly naked singularties}.

The mass parameter $M=0$ in the  Janis-Newman-Winicour solution describes the situation of a purely scalar field.
We do not consider this case  henceforth in this paper.


\section{\label{sec:TD&MC} Time Delay, Magnification Centroid, and Total Magnification}

We consider light propagation  in a static spherically symmetric   
spacetime described by the line element given by Eq. $(\ref{GenMetric})$. The spherical symmetry of the spacetime allows
us to consider, without loss of generality, null geodesics in the equatorial plane. We first 
obtain time required for light to travel from a source at coordinates $\{r,\vartheta=\pi/2,
\varphi=\varphi_1\}$
to the closest distance of approach at coordinates $\{r_0,\vartheta=\pi/2,\varphi=\varphi_2\}$.
Following the method used in \cite{Wein72}, a straightforward calculation thus gives the time required for light to travel from
$r$ to $r_0$ (or $r_0$ to $r$) that is expressed by
\begin{equation}
  t\lt(r,r_0\rt)=t\lt(r_0,r\rt)={{\int_{r_0}}^r} 
    \sqrt{\frac{A\lt(r\rt)/B\lt(r\rt)} 
               {1-\lt(\frac{r_0}{r}\rt)^2 \frac{B\lt(r\rt)}{B\lt(r_0\rt)} \frac{D\lt(r_0\rt)}{D\lt(r\rt)}}} ~ ~ dr  \text{.}             
    \label{trr0}
\end{equation}
$D\lt(r\rt)=1$ in the above equation readily gives the result in \cite{Wein72}. Let ${\cal R}_s$ and ${\cal R}_o$  
denote, respectively, the radial coordinates of the source and the observer measured from the center of mass of the deflector
(lens). We now express these distances in terms of the constant  parameter $b$ (in the Janis-Newman-Winicour metric) by introducing
\begin{equation}
  {\cal X}_s = \frac{{\cal R}_s}{b} ~ ~ ~ ~ \text{and} ~ ~ ~ ~  {\cal X}_o = \frac{{\cal R}_o}{b} \text{.}
  \label{XtoR}
\end{equation}
The time delay ${\tau}\lt(\rho_0\rt)$ of light traveling from the source to the observer with the closest
of approach $\rho_0$ is defined as the difference between the light travel time for the actual ray in the 
gravitational field  of the lens (deflector) and the travel time for the straight path between the source and the observer in the
absence of the lens (i.e., if there were no  gravitational fields.)  As mentioned in Section \ref{sec:LE&DE}, we  do not
consider the case of purely scalar field  in this paper; therefore, we assume that $\nu \neq 0$ (i.e., $M \neq 0$).
Using equations  $(\ref{Janis-Newman-WinicourMetric})$, $(\ref{nub})$ and $(\ref{trr0})$, and the geometry of the lens diagram
(see Fig. 1 in \cite{VE00}), we obtain the following expression for the time delay in the Janis-Newman-Winicour spacetime:
\begin{equation}
  \tau\lt(\rho_0\rt) 
      = \frac{2M}{\nu} 
         \lt[         
             {\int_{\rho_0}}^{{\cal X}_s} \frac{d\rho}{f\lt(\rho\rt)} + {\int_{\rho_0}}^{{\cal X}_o} \frac{d\rho}{f\lt(\rho\rt)}          
          \rt]  
       - D_s \sec\beta                                        
    \label{tau}
\end{equation}
with
\begin{eqnarray}
    {\cal X}_s &=& \frac{\nu}{2} \frac{D_s}{M} \sqrt{   \lt(\frac{D_{ds}}{D_s}\rt)^2+\tan^2\beta  } \text{,}\nonumber\\
    {\cal X}_o &=& \frac{\nu}{2} \frac{D_d}{M} \text{,} 
\end{eqnarray}
and
\begin{equation}
    f\lt(\rho\rt)={ \sqrt{ \lt(1-\frac{1}{\rho}\rt)^{2\nu}-\lt(\frac{\rho_0}{\rho}\rt)^2\lt(1-\frac{1}{\rho}\rt)^{4\nu-1}\lt(1-\frac{1}{\rho_0}\rt)^{1-2\nu}}}  \text{.}                             
    \label{f}
\end{equation}
The first and second terms in Eq. $(\ref{tau})$ give, respectively, the travel time of the light from the source to the point of
closest approach and from that point  to the observer. The last term gives the light travel time from the source to the observer in
the absence of any gravitational field.


We use the equation  $( \ref{tau})$   for computations in the next section. However, to see the behavior of the time delay 
function for a light ray traveling in the weak gravitation field far away from the  lens, we carry out some analytical 
calculations following the method used in \cite{KP05}. We obtain the time delay for images with large impact parameters. For 
given angular positions of the source and image, the time
delay is given by
\begin{equation}
    \tau(\theta, \beta) = \frac{1}{2} \frac{D_d D_s}{D_{ds}}
    \left[ \left(2-\frac{1}{\nu}\right)\theta_E^2 + \beta^2 - \theta^2
    - \ln \left( \frac{\theta^2 D_d}{4 D_{ds}} \right) \right] \text{,}
\end{equation}
where
\begin{equation}
    \theta_E = \sqrt{4 M \ \frac{D_{ds}}{D_d D_s}}
\end{equation}
is an approximate expression for angular radius of  the Einstein ring of  Schwarzschild lensing.

For the direct image, $\beta^2-\theta^2$ decreases with an increase in $|\beta|$.  Therefore, for $\nu<1/2$ (SNS), 
the time delay of the direct image is negative for
large $\beta$. This fact reflects in our computations in the next section.


We denote the time delay in the outermost image on the same side as the source (also called the direct image)
by $\tau_{os}$.  The {\em differential time delay} $\Delta\tau$ of an image with time delay $\tau$  is defined by 
\begin{equation}
   \Delta\tau = \tau - \tau_{os} \text{.}
\end{equation}
(When there is only one image on the same side as the source, we use the symbol $\tau_{s}$ instead of $\tau_{os}$.)
The differential time delay is thus measured in reference  to the direct  image.
The {\em magnification-weighted centroid position} (also called  {\em magnification centroid})  of images
is defined by 
\begin{equation}
  \hat{\Theta} = \frac{\sum \theta_i |\mu_i|}{\sum |\mu_i|} \text{.}
\end{equation}
Angles measured in clockwise and anti-clockwise directions from the optical axis have positive and negative signs,
respectively. Further, the {\em magnification centroid shift}  of images is defined by
\begin{equation}
 \Delta\hat{\Theta} =  \beta - \hat{\Theta} \text{.}
\end{equation}
The {\em total absolute magnification} (also, simply called {\em total magnification}) $\mu_{tot}$ is defined  by
\begin{equation}
 \mu_{tot}  = \sum |\mu_i| \text{.}
\end{equation}
The magnification centroid and the total magnification are very important physical quantities in studying microlensing 
when the images are not resolved.


\section{\label{sec:Comp} Computations}

Virbhadra and Ellis \cite{VE02}   modeled massive dark objects (MDOs) at the centers of four different galaxies (including 
our galaxy) as   Schwarzschild black holes (SBH) and Janis-Newman-Winicour naked singularities, and studied  
point source gravitational lensing by them. They obtained the angular positions of critical curves and caustics,
and studied  the variation of magnification  against the angular position of images near the critical curves.
However, they did not study the time delay, magnification centroid, and the total  magnification; we accomplish these
tasks in this paper.

We now consider the MDO at the center of our galaxy with the recent values for the mass $M=3.61\times 10^6  M_{\odot}$ 
and the lens-observer distance $D_d= 7.62$ kpc \cite{Eisenetal05}. As in \cite{VE02}, we take
the lens (MDO) to be situated half way between the source and the observer, i.e., $D_{ds}/D_s=1/2$. We model the Galactic
MDO as  a Schwarzschild black hole as well as  Janis-Newman-Winicour WNS, MSNS, and SNS lenses.  As we are considering 
updated values for 
$M$ and $D_d$ of the Galactic MDO, we first re-compute critical curves and caustics, and their corresponding deflection 
angles of the light ray 
for several values of $\nu$. For continuity and clarity in the analysis of the results, we will also mention some 
results from \cite{VNC98,VE02} in the present and  next sections of this paper.

The existence of a photon sphere
covering a gravitational lens is a sufficient (but not necessary) condition for the occurrence of relativistic images.
The SBH as well as WNS both are contained inside a photon sphere and both give rise to relativistic images.
The MSNS lensing also produces  relativistic images, even though it is not covered by a photon sphere.
It is known   that 
relativistic images are transient and extremely demagnified, and  therefore their observations do not seem to be feasible 
in the near future \cite{VE00}. Hence, we do not do  computations for  relativistic  images in this paper.

\begingroup
\squeezetable
\begin{table*}
\caption{\label{tab:Table1}  Critical curves and caustics due to gravitational lensing by the Galactic  MDO modeled as the 
                             Schwarzschild black hole; weakly, marginally strongly, and strongly naked singularities.
          $\theta_E$, $\theta_r$, and $\beta_r$ denote respectively the angular positions of the tangential
             critical curves (Einstein rings), radial critical curves, and radial caustics, whereas $\hat{\alpha}$
             stands for the corresponding light deflection angles.
   $({\rm a})$ The lens has mass  $M= 3.61 \times 10^6 M_{\odot} $ and the distance $D_d =  7.62$ kpc  so that 
              $M/D_d \approx 2.26 \times  10^{-11} $). The ratio of the source-lens distance to the source-observer 
              distance, i.e.,  $D_{ds}/D_s=1/2$.  All angular positions  are given  in arcseconds.
        }
\begin{ruledtabular}
  \begin{tabular}{l|cc|ccc|cc}
\multicolumn{1}{c|}{$\nu$}&
\multicolumn{2}{c|}{Inner Einstein ring}&
\multicolumn{3}{c|}{Radial critical curve and caustic}&
\multicolumn{2}{c}{Outer Einstein ring} \\
& $\theta_E$    &    $\hat{\alpha}$    &         $\theta_r$    &    $\hat{\alpha}$      &   $\beta_r$  & $\theta_E$    &    $\hat{\alpha}$      \\
\hline
$1.0   $&$ \times   $&$ \times    $&$ \times    $&$ \times   $&$ \times    $&$ 1.388176 $&$ 2.776352 $\\
 $0.9   $&$ \times   $&$ \times   $&$ \times    $&$ \times   $&$ \times    $&$ 1.388176 $&$ 2.776352 $\\
 $0.8   $&$ \times   $&$ \times   $&$ \times    $&$ \times   $&$ \times    $&$ 1.388176 $&$ 2.776352 $\\
 $0.7   $&$ \times   $&$ \times   $&$ \times    $&$ \times   $&$ \times    $&$ 1.388176 $&$ 2.776351 $\\
 $0.6   $&$ \times   $&$ \times   $&$ \times    $&$ \times   $&$ \times    $&$ 1.388175 $&$ 2.776351 $\\
 $0.5   $&$ \times   $&$ \times   $&$ \times    $&$ \times   $&$ \times    $&$ 1.388175 $&$ 2.776350 $\\
 $0.4   $&$ 0.000012 $&$ 0.000024 $&$ -0.000019 $&$ 286883.8 $&$ 252026.7  $&$ 1.388174 $&$ 2.776348 $\\
 $0.3   $&$ 0.000015 $&$ 0.000030 $&$ -0.000026 $&$ 121357.9 $&$ 66413.69  $&$ 1.388172 $&$ 2.776343 $\\
 $0.2   $&$ 0.000027 $&$ 0.000054 $&$ -0.000051 $&$ 46469.19 $&$ 23533.19  $&$ 1.388165 $&$ 2.776330 $\\
 $0.1   $&$ 0.000095 $&$ 0.000191 $&$ -0.000188 $&$ 10759.36 $&$ 5383.343  $&$ 1.388131 $&$ 2.776262 $\\
 $0.05  $&$ 0.000371 $&$ 0.000741 $&$ -0.000739 $&$ 2641.816 $&$ 1320.961  $&$ 1.387993 $&$ 2.775986 $\\
 $0.04  $&$ 0.000577 $&$ 0.001154 $&$ -0.001151 $&$ 1687.159 $&$ 843.5924  $&$ 1.387890 $&$ 2.775780 $\\
 $0.03  $&$ 0.001023 $&$ 0.002046 $&$ -0.002043 $&$ 947.4574 $&$ 473.7291  $&$ 1.387667 $&$ 2.775334 $\\
 $0.02  $&$ 0.002297 $&$ 0.004593 $&$ -0.004591 $&$ 420.5955 $&$ 210.2934  $&$ 1.387029 $&$ 2.774057 $\\
 $0.01  $&$ 0.009176 $&$ 0.018352 $&$ -0.018345 $&$ 105.0746 $&$ 52.51893  $&$ 1.383568 $&$ 2.767135 $\\
 $0.005 $&$ 0.036717 $&$ 0.073434 $&$ -0.073177 $&$ 26.26379 $&$ 13.05872  $&$ 1.369456 $&$ 2.738911 $\\
 $0.004 $&$ 0.057426 $&$ 0.114853 $&$ -0.113887 $&$ 16.80784 $&$ 8.290033  $&$ 1.358574 $&$ 2.717148 $\\
 $0.003 $&$ 0.102472 $&$ 0.204944 $&$ -0.199693 $&$ 9.450634 $&$ 4.525624  $&$ 1.334103 $&$ 2.668206 $\\
 $0.002 $&$ 0.236134 $&$ 0.472268 $&$ -0.420122 $&$ 4.166782 $&$ 1.663269  $&$ 1.254966 $&$ 2.509932 $\\
 $0.001 $&$ \times   $&$ \times   $&$ 1.115015  $&$ 0.613237 $&$  0.808397 $&$ \times   $&$ \times$  \\
\end{tabular}
\end{ruledtabular}
\end{table*}
\endgroup
\begingroup
\squeezetable
\begin{table*}
\caption{\label{tab:Table2}  Image positions due to lensing by the  Galactic MDO 
            modeled as the Schwarzschild black hole ($\nu =1$),  and their respective bending angles, magnifications,
            time delays, and differential time delays.
   $({\rm a})$ $\theta$ and  $\hat{\alpha}$ respectively stand for the angular positions of images and their
             corresponding deflection angles. $\mu$, $\tau$, and $\Delta\tau$ represent the magnification,
             time delay, and differential time delay of the images, respectively. 
   $({\rm b})$ The subscripts $s$ and $o$ on the symbols  respectively denote for the images on the same and opposite side of the source.
   $({\rm c})$  The same as  (${\rm a}$) of Table I.
       } 
\begin{ruledtabular}
  \begin{tabular}{l|ccccc|cccc}  
\multicolumn{1}{c|}{$\beta$}&
\multicolumn{5}{c|}{Image on the opposite side from the source}&
\multicolumn{4}{c}{Image on the same side as the source}\\
&    $\theta_o$            &         $\hat{\alpha}_o$ &    $\mu_o$                 &   $\tau_o$     & $\Delta\tau_o$  &    $\theta_s$           &  $\hat{\alpha}_s$ &    $\mu_s$ & $\tau_s$    \\
\hline
 $ 0       $&$ -1.388176 $&$ 2.776352 $&$ \times    $&$ 14.92209 $&$  0        $&$ 1.388176 $&$ 2.776352 $&$ \times   $&$ 14.92209 $\\
 $ 10^{-6} $&$ -1.388176 $&$ 2.776353 $&$ -694084.2 $&$ 14.92209 $&$  0.000002 $&$ 1.388177 $&$ 2.776351 $&$ 694085.2 $&$ 14.92209 $\\ 
 $ 10^{-5} $&$ -1.388171 $&$ 2.776362 $&$ -69407.97 $&$ 14.92210 $&$  0.000017 $&$ 1.388181 $&$ 2.776342 $&$ 69408.97 $&$ 14.92208 $\\
 $ 10^{-4} $&$ -1.388126 $&$ 2.776452 $&$ -6940.347 $&$ 14.92218 $&$  0.000171 $&$ 1.388226 $&$ 2.776252 $&$ 6941.347 $&$ 14.92201 $\\
 $ 10^{-3} $&$ -1.387676 $&$ 2.777353 $&$ -693.5848 $&$ 14.92294 $&$  0.001706 $&$ 1.388676 $&$ 2.775353 $&$ 694.5848 $&$ 14.92124 $\\
 $ 10^{-2} $&$ -1.383185 $&$ 2.786370 $&$ -68.90982 $&$ 14.93064 $&$  0.017060 $&$ 1.393185 $&$ 2.766370 $&$ 69.90982 $&$ 14.91358 $\\
 $ 10^{-1} $&$ -1.339077 $&$ 2.878153 $&$ -6.454348 $&$ 15.00895 $&$  0.170636 $&$ 1.439076 $&$ 2.678152 $&$ 7.454345 $&$ 14.83831 $\\
 $ 1       $&$ -0.975480 $&$ 3.950960 $&$ -0.322455 $&$ 15.94681 $&$  1.742193 $&$ 1.975475 $&$ 1.950951 $&$ 1.322453 $&$ 14.20461 $\\
 $ 2       $&$ -0.710863 $&$ 5.421726 $&$ -0.073840 $&$ 17.38033 $&$  3.687537 $&$ 2.710855 $&$ 1.421709 $&$ 1.073838 $&$ 13.69280 $\\
 $ 3       $&$ -0.543786 $&$ 7.087573 $&$ -0.024114 $&$ 19.29818 $&$  5.987040 $&$ 3.543776 $&$ 1.087553 $&$ 1.024113 $&$ 13.31114 $\\
 $ 4       $&$ -0.434559 $&$ 8.869117 $&$ -0.009696 $&$ 21.74718 $&$  8.734391 $&$ 4.434547 $&$ 0.869094 $&$ 1.009695 $&$ 13.01279 $\\
 $ 5       $&$ -0.359561 $&$ 10.71912 $&$ -0.004521 $&$ 24.75495 $&$  11.98479 $&$ 5.359549 $&$ 0.719098 $&$ 1.004521 $&$ 12.77016 $\\
 $ 6       $&$ -0.305617 $&$ 12.61123 $&$ -0.002355 $&$ 28.33806 $&$  15.77141 $&$ 6.305604 $&$ 0.611208 $&$ 1.002354 $&$ 12.56665 $\\
 $ 7       $&$ -0.265251 $&$ 14.53050 $&$ -0.001335 $&$ 32.50696 $&$  20.11513 $&$ 7.265238 $&$ 0.530476 $&$ 1.001335 $&$ 12.39183 $\\
 $ 8       $&$ -0.234044 $&$ 16.46809 $&$ -0.000809 $&$ 37.26852 $&$  25.02969 $&$ 8.234031 $&$ 0.468062 $&$ 1.000808 $&$ 12.23882 $\\
 $ 9       $&$ -0.209261 $&$ 18.41852 $&$ -0.000517 $&$ 42.62746 $&$  30.52455 $&$ 9.209248 $&$ 0.418496 $&$ 1.000517 $&$ 12.10291 $\\
 $ 10      $&$ -0.189138 $&$ 20.37828 $&$ -0.000345 $&$ 48.58714 $&$  36.60642 $&$ 10.18912 $&$ 0.378250 $&$ 1.000345 $&$ 11.98072 $\\
 \end{tabular}
\end{ruledtabular}
 \end{table*}
 \endgroup
\begingroup
\squeezetable
\begin{table*}
\caption{\label{tab:Table3} Image positions due to lensing by the  Galactic MDO 
            modeled as the weakly naked singularity ($\nu =0.7$), and their respective bending angles, magnifications,
            time delays and differential time delays.
       ({\rm a})  The same as (${\rm a}$)  and (${\rm b}$)  of Table II.  (${\rm b}$) The same as  (${\rm a}$)  of Table I.
       }
\begin{ruledtabular}
\begin{tabular}{l|ccccc|cccc}
\multicolumn{1}{c|}{$\beta$}&
\multicolumn{5}{c|}{Image on the opposite side from the source}&
\multicolumn{4}{c}{Image on the same side as the source}\\
&    $\theta_o$            &         $\hat{\alpha}_o$ &    $\mu_o$     &   $\tau_o$     & $\Delta\tau_o$  &    $\theta_s$    &  $\hat{\alpha}_s$ &    $\mu_s$   &   $\tau_s$    \\
\hline
$0       $&$ -1.388176 $&$ 2.776351 $&$ \times    $&$ 14.66836 $&$  0       $&$ 1.388176 $&$ 2.776351 $&$ \times   $&$ 14.66836 $\\
$10^{-6} $&$ -1.388175 $&$ 2.776352 $&$ -694084.2 $&$ 14.66836 $&$ 0.000002 $&$ 1.388176 $&$ 2.776350 $&$ 694085.2 $&$ 14.66835 $\\
$10^{-5} $&$ -1.388171 $&$ 2.776361 $&$ -69407.97 $&$ 14.66836 $&$ 0.000017 $&$ 1.388181 $&$ 2.776341 $&$ 69408.97 $&$ 14.66835 $\\
$10^{-4} $&$ -1.388126 $&$ 2.776451 $&$ -6940.347 $&$ 14.66844 $&$ 0.000171 $&$ 1.388226 $&$ 2.776251 $&$ 6941.347 $&$ 14.66827 $\\
$10^{-3} $&$ -1.387676 $&$ 2.777352 $&$ -693.5848 $&$ 14.66921 $&$ 0.001706 $&$ 1.388676 $&$ 2.775352 $&$ 694.5848 $&$ 14.66750 $\\
$10^{-2} $&$ -1.383185 $&$ 2.786370 $&$ -68.90982 $&$ 14.67690 $&$ 0.017060 $&$ 1.393185 $&$ 2.766369 $&$ 69.90982 $&$ 14.65984 $\\ 
$10^{-1} $&$ -1.339076 $&$ 2.878152 $&$ -6.454347 $&$ 14.75521 $&$ 0.170636 $&$ 1.439076 $&$ 2.678151 $&$ 7.454345 $&$ 14.58457 $\\
$1       $&$ -0.975479 $&$ 3.950959 $&$ -0.322455 $&$ 15.69307 $&$ 1.742192 $&$ 1.975475 $&$ 1.950950 $&$ 1.322453 $&$ 13.95088 $\\
$2       $&$ -0.710862 $&$ 5.421724 $&$ -0.073840 $&$ 17.12660 $&$ 3.687536 $&$ 2.710855 $&$ 1.421709 $&$ 1.073838 $&$ 13.43906 $\\
$3       $&$ -0.543786 $&$ 7.087571 $&$ -0.024113 $&$ 19.04444 $&$ 5.987039 $&$ 3.543776 $&$ 1.087552 $&$ 1.024113 $&$ 13.05740 $\\
$4       $&$ -0.434558 $&$ 8.869115 $&$ -0.009696 $&$ 21.49345 $&$ 8.734389 $&$ 4.434547 $&$ 0.869094 $&$ 1.009695 $&$ 12.75906 $\\
$5       $&$ -0.359560 $&$ 10.71912 $&$ -0.004521 $&$ 24.50121 $&$ 11.98478 $&$ 5.359549 $&$ 0.719097 $&$ 1.004521 $&$ 12.51643 $\\
$6       $&$ -0.305616 $&$ 12.61123 $&$ -0.002355 $&$ 28.08432 $&$ 15.77141 $&$ 6.305604 $&$ 0.611208 $&$ 1.002354 $&$ 12.31292 $\\
$7       $&$ -0.265250 $&$ 14.53050 $&$ -0.001335 $&$ 32.25322 $&$ 20.11513 $&$ 7.265238 $&$ 0.530476 $&$ 1.001335 $&$ 12.13810 $\\
$8       $&$ -0.234043 $&$ 16.46809 $&$ -0.000809 $&$ 37.01478 $&$ 25.02969 $&$ 8.234031 $&$ 0.468062 $&$ 1.000808 $&$ 11.98509 $\\
$9       $&$ -0.209260 $&$ 18.41852 $&$ -0.000517 $&$ 42.37372 $&$ 30.52454 $&$ 9.209248 $&$ 0.418496 $&$ 1.000517 $&$ 11.84918 $\\
$10      $&$ -0.189137 $&$ 20.37827 $&$ -0.000345 $&$ 48.33340 $&$ 36.60642 $&$ 10.18912 $&$ 0.378250 $&$ 1.000345 $&$ 11.72698$\\ 
 \end{tabular}
\end{ruledtabular}
 \end{table*}
 \endgroup

\begingroup
\squeezetable
\begin{table*}
\caption{\label{tab:Table4} Image positions due to lensing by the Galactic  MDO 
            modeled as the marginally strongly naked singularity ($\nu =0.5$) and their respective bending angles, magnifications,
            time delays and differential time delays.
        ({\rm a})  The same as (${\rm a}$)  and (${\rm b}$) of Table II.   (${\rm b}$) The same as (${\rm a}$) of Table I.
      }
\begin{ruledtabular}
  \begin{tabular}{l|ccccc|cccc}
\multicolumn{1}{c|}{$\beta$}&
\multicolumn{5}{c|}{Image on the opposite side from the source}&
\multicolumn{4}{c}{Image on the same side as the source}\\
&    $\theta_o$            &         $\hat{\alpha}_o$ &    $\mu_o$                 &   $\tau_o$     & $\Delta\tau_o$  &    $\theta_s$           &  $\hat{\alpha}_s$ &    $\mu_s$ & $\tau_s$    \\
\hline
 $0        $&$ -1.388175 $&$ 2.776350 $&$ \times    $&$ 14.33004 $&$ 0        $&$ 1.388175 $&$ 2.776350 $&$ \times   $&$ 14.33004 $ \\
 $10^{-6}  $&$ -1.388174 $&$ 2.776351 $&$ -694084.2 $&$ 14.33004 $&$ 0.000002 $&$ 1.388175 $&$ 2.776349 $&$ 694085.2 $&$ 14.33004 $ \\
 $10^{-5}  $&$ -1.388170 $&$ 2.776360 $&$ -69407.97 $&$ 14.33005 $&$ 0.000017 $&$ 1.388180 $&$ 2.776340 $&$ 69408.97 $&$ 14.33003 $ \\
 $10^{-4}  $&$ -1.388125 $&$ 2.776450 $&$ -6940.347 $&$ 14.33013 $&$ 0.000171 $&$ 1.388225 $&$ 2.776250 $&$ 6941.347 $&$ 14.32995 $ \\
 $10^{-3}  $&$ -1.387675 $&$ 2.777350 $&$ -693.5848 $&$ 14.33089 $&$ 0.001706 $&$ 1.388675 $&$ 2.775350 $&$ 694.5848 $&$ 14.32919 $ \\
 $10^{-2}  $&$ -1.383184 $&$ 2.786368 $&$ -68.90982 $&$ 14.33859 $&$ 0.017060 $&$ 1.393184 $&$ 2.766368 $&$ 69.90982 $&$ 14.32153 $ \\
 $10^{-1}  $&$ -1.339075 $&$ 2.878150 $&$ -6.454347 $&$ 14.41689 $&$ 0.170636 $&$ 1.439075 $&$ 2.678150 $&$ 7.454345 $&$ 14.24626 $ \\
 $1        $&$ -0.975478 $&$ 3.950956 $&$ -0.322455 $&$ 15.35475 $&$ 1.742191 $&$ 1.975474 $&$ 1.950949 $&$ 1.322453 $&$ 13.61256 $ \\
 $2        $&$ -0.710861 $&$ 5.421721 $&$ -0.073840 $&$ 16.78828 $&$ 3.687533 $&$ 2.710854 $&$ 1.421708 $&$ 1.073838 $&$ 13.10075 $ \\
 $3        $&$ -0.543784 $&$ 7.087568 $&$ -0.024113 $&$ 18.70612 $&$ 5.987035 $&$ 3.543776 $&$ 1.087552 $&$ 1.024113 $&$ 12.71909 $ \\
 $4        $&$ -0.434556 $&$ 8.869112 $&$ -0.009696 $&$ 21.15513 $&$ 8.734385 $&$ 4.434547 $&$ 0.869094 $&$ 1.009695 $&$ 12.42074 $ \\
 $5        $&$ -0.359558 $&$ 10.71912 $&$ -0.004521 $&$ 24.16289 $&$ 11.98478 $&$ 5.359549 $&$ 0.719097 $&$ 1.004521 $&$ 12.17811 $ \\
 $6        $&$ -0.305614 $&$ 12.61123 $&$ -0.002354 $&$ 27.74600 $&$ 15.77140 $&$ 6.305604 $&$ 0.611208 $&$ 1.002354 $&$ 11.97460 $ \\
 $7        $&$ -0.265248 $&$ 14.53050 $&$ -0.001335 $&$ 31.91490 $&$ 20.11512 $&$ 7.265238 $&$ 0.530476 $&$ 1.001335 $&$ 11.79978 $ \\
 $8        $&$ -0.234041 $&$ 16.46808 $&$ -0.000809 $&$ 36.67646 $&$ 25.02968 $&$ 8.234031 $&$ 0.468062 $&$ 1.000808 $&$ 11.64678 $ \\
 $9        $&$ -0.209259 $&$ 18.41852 $&$ -0.000517 $&$ 42.03539 $&$ 30.52453 $&$ 9.209248 $&$ 0.418496 $&$ 1.000517 $&$ 11.51086 $ \\
 $10       $&$ -0.189135 $&$ 20.37827 $&$ -0.000345 $&$ 47.99508 $&$ 36.60640 $&$ 10.18912 $&$ 0.378250 $&$ 1.000345 $&$ 11.38867 $ \\
 \end{tabular}
\end{ruledtabular}
 \end{table*}
 \endgroup
\begingroup
\squeezetable
\begin{table*}
\caption{\label{tab:Table5} Image positions  on the same side as the source due to lensing by the Galactic MDO 
            modeled as a strongly naked singularity ($\nu =0.04$), and their respective bending angles, magnifications, time delays
          and differential time delays.
       $({\rm a})$ The same as (${\rm a}$) of Table II.
       $({\rm b})$ The subscripts $is$ and $os$ on the symbols  respectively denote for the inner and outer images  on the same side as the source.
       $({\rm c})$ The same as  (${\rm a}$) of Table I. }
\begin{ruledtabular}
  \begin{tabular}{l|ccccc|cccc}
\multicolumn{1}{c|}{$\beta$}&
\multicolumn{5}{c|}{Inner image on the same side as the source}&
\multicolumn{4}{c}{Outer image on the same side as the source}\\
&    $\theta_{is}$  &   $\hat{\alpha}_{is}$ &    $\mu_{is}$   &   $\tau_{is}$  & $\Delta\tau_{is}$  &    $\theta_{os}$  &  $\hat{\alpha}_{os}$ &  $\mu_{os}$ & $\tau_{os}$    \\
\hline
 
$10^{-6} $&$ 0.000577 $&$  0.001152 $&$ -0.000099            $&$ 8.164602 $&$ 7.452654 $&$ 1.387890 $&$ 2.775779 $&$ 694085.2 $&$  0.711948 $ \\
  $10^{-5} $&$ 0.000577 $&$  0.001134 $&$ -9.9 \times 10^{-6}  $&$ 8.164602 $&$ 7.452662 $&$ 1.387895 $&$ 2.775770 $&$ 69408.97 $&$  0.711940 $ \\
  $10^{-4} $&$ 0.000577 $&$  0.000954 $&$ -9.9 \times 10^{-7}  $&$ 8.164602 $&$ 7.452738 $&$ 1.387940 $&$ 2.775680 $&$ 6941.347 $&$  0.711863 $ \\
  $10^{-3} $&$ 0.000577 $&$ -0.000846 $&$ -9.9 \times 10^{-8}  $&$ 8.164602 $&$ 7.453505 $&$ 1.388390 $&$ 2.774780 $&$ 694.5849 $&$  0.711097 $ \\
  $10^{-2} $&$ 0.000577 $&$ -0.018846 $&$ -9.9 \times 10^{-9}  $&$ 8.164629 $&$ 7.461187 $&$ 1.392900 $&$ 2.765800 $&$ 69.90987 $&$  0.703442 $ \\
  $10^{-1} $&$ 0.000577 $&$ -0.198846 $&$ -9.9 \times 10^{-10} $&$ 8.167637 $&$ 7.539395 $&$ 1.438800 $&$ 2.677600 $&$ 7.454397 $&$  0.628242 $ \\
  $1       $&$ 0.000577 $&$ -1.998846 $&$ -9.9 \times 10^{-11} $&$ 8.471271 $&$ 8.476159 $&$ 1.975286 $&$ 1.950572 $&$ 1.322496 $&$ -0.004887 $ \\
  $2       $&$ 0.000577 $&$ -3.998847 $&$ -4.9 \times 10^{-11} $&$ 9.391988 $&$ 9.908244 $&$ 2.710736 $&$ 1.421472 $&$ 1.073866 $&$ -0.516256 $ \\
  $3       $&$ 0.000576 $&$ -5.998847 $&$ -3.3 \times 10^{-11} $&$ 10.92675 $&$ 11.82434 $&$ 3.543700 $&$ 1.087400 $&$ 1.024129 $&$ -0.897592 $ \\
  $4       $&$ 0.000576 $&$ -7.998848 $&$ -2.5 \times 10^{-11} $&$ 13.07556 $&$ 14.27126 $&$ 4.434496 $&$ 0.868992 $&$ 1.009705 $&$ -1.195693 $ \\
  $5       $&$ 0.000576 $&$ -9.998848 $&$ -2.0 \times 10^{-11} $&$ 15.83842 $&$ 17.27655 $&$ 5.359513 $&$ 0.719026 $&$ 1.004527 $&$ -1.438130 $ \\
  $6       $&$ 0.000576 $&$ -11.99885 $&$ -1.6 \times 10^{-11} $&$ 19.21533 $&$ 20.85681 $&$ 6.305578 $&$ 0.611155 $&$ 1.002358 $&$ -1.641480 $ \\
  $7       $&$ 0.000576 $&$ -13.99885 $&$ -1.4 \times 10^{-11} $&$ 23.20628 $&$ 25.02245 $&$ 7.265218 $&$ 0.530436 $&$ 1.001337 $&$ -1.816167 $ \\
  $8       $&$ 0.000576 $&$ -15.99885 $&$ -1.2 \times 10^{-11} $&$ 27.81128 $&$ 29.78034 $&$ 8.234015 $&$ 0.468030 $&$ 1.000810 $&$ -1.969056 $ \\
  $9       $&$ 0.000575 $&$ -17.99885 $&$ -1.1 \times 10^{-11} $&$ 33.03033 $&$ 35.13520 $&$ 9.209235 $&$ 0.418471 $&$ 1.000518 $&$ -2.104868 $ \\
  $10      $&$ 0.000575 $&$ -19.99885 $&$ -9.7 \times 10^{-12} $&$ 38.86343 $&$ 41.09040 $&$ 10.18911 $&$ 0.378229 $&$ 1.000346 $&$ -2.226966 $ \\
 \end{tabular}
\end{ruledtabular}
 \end{table*}
 \endgroup

\begingroup
\squeezetable
\begin{table*}
\caption{\label{tab:Table6} Image positions  on the opposite  side of the source due to lensing by the Galactic MDO 
            modeled as a strongly naked singularity ($\nu =0.04$), and their respective bending angles, magnifications, time delays 
           and differential time delays.
       $({\rm a})$ The same as (${\rm a}$) of Table II.
       $({\rm b})$ The subscripts $io$ and $oo$ on the symbols  respectively denote for the inner and outer images  on the opposite side from the source.
       $({\rm c})$ The same as  (${\rm a}$) of Table I.
   }
\begin{ruledtabular}
  \begin{tabular}{l|ccccc|ccccc}
\multicolumn{1}{c|}{$\beta$}&
\multicolumn{5}{c|}{Outer image on the opposite side from the source}&
\multicolumn{5}{c}{Inner image on the opposite side from the source}\\
&    $\theta_{oo}$  & $\hat{\alpha}_{oo}$ & $\mu_{oo}$ &   $\tau_{oo}$ & $\Delta\tau_{oo}$    &    $\theta_{io}$  &  $\hat{\alpha}_{io}$ & $\mu_{io}$ & $\tau_{io}$ & $\Delta\tau_{io}$  \\
\hline
  $10^{-6} $&$ -1.387889 $&$ 2.775781 $&$ -694084.2 $&$ 0.711950 $&$ 0.000002            $&$ -0.000577 $&$ 0.001156 $&$ 0.000099             $&$ 8.164602 $&$ 7.452654 $ \\
  $10^{-5} $&$ -1.387885 $&$ 2.775790 $&$ -69407.97 $&$ 0.711957 $&$ 0.000017            $&$ -0.000577 $&$ 0.001174 $&$ 9.9 \times 10^{-6}   $&$ 8.164602 $&$ 7.452662 $ \\
  $10^{-4} $&$ -1.387840 $&$ 2.775880 $&$ -6940.347 $&$ 0.712034 $&$ 0.000170            $&$ -0.000577 $&$ 0.001354 $&$ 9.9 \times 10^{-7}   $&$ 8.164602 $&$ 7.452738 $ \\
  $10^{-3} $&$ -1.387390 $&$ 2.776780 $&$ -693.5848 $&$ 0.712801 $&$ 0.001704            $&$ -0.000577 $&$ 0.003154 $&$ 9.9 \times 10^{-8}   $&$ 8.164603 $&$ 7.453506 $ \\
  $10^{-2} $&$ -1.382898 $&$ 2.785796 $&$ -68.90977 $&$ 0.720486 $&$ 0.017045            $&$ -0.000577 $&$ 0.021154 $&$ 9.9 \times 10^{-9}   $&$ 8.164636 $&$ 7.461194 $ \\
  $10^{-1} $&$ -1.338780 $&$ 2.877560 $&$ -6.454297 $&$ 0.798725 $&$ 0.170483            $&$ -0.000577 $&$ 0.201154 $&$ 9.9 \times 10^{-10}  $&$ 8.167708 $&$ 7.539465 $ \\
  $1       $&$ -0.975097 $&$ 3.950193 $&$ -0.322412 $&$ 1.735749 $&$ 1.740636            $&$ -0.000577 $&$ 2.001154 $&$ 9.9 \times 10^{-11}  $&$ 8.471980 $&$ 8.476867 $ \\
  $2       $&$ -0.710409 $&$ 5.420818 $&$ -0.073812 $&$ 3.168027 $&$ 3.684283            $&$ -0.000577 $&$ 4.001155 $&$ 5.0 \times 10^{-11}  $&$ 9.393405 $&$ 9.909661 $ \\
  $3       $&$ -0.543290 $&$ 7.086579 $&$ -0.024097 $&$ 5.084252 $&$ 5.981844            $&$ -0.000577 $&$ 6.001155 $&$ 3.3 \times 10^{-11}  $&$ 10.92888 $&$ 11.82647 $ \\
  $4       $&$ -0.434037 $&$ 8.868073 $&$ -0.009686 $&$ 7.531250 $&$ 8.726943            $&$ -0.000578 $&$ 8.001155 $&$ 2.5 \times 10^{-11}  $&$ 13.07840 $&$ 14.27409 $ \\
  $5       $&$ -0.359024 $&$ 10.71805 $&$ -0.004515 $&$ 10.53661 $&$ 11.97474            $&$ -0.000578 $&$ 10.00116 $&$ 2.0 \times 10^{-11}  $&$ 15.84197 $&$ 17.28010 $ \\
  $6       $&$ -0.305070 $&$ 12.61014 $&$ -0.002351 $&$ 14.11691 $&$ 15.75839            $&$ -0.000578 $&$ 12.00116 $&$ 1.7 \times 10^{-11}  $&$ 19.21958 $&$ 20.86106 $ \\
  $7       $&$ -0.264698 $&$ 14.52939 $&$ -0.001332 $&$ 18.28259 $&$ 20.09876            $&$ -0.000578 $&$ 14.00116 $&$ 1.4 \times 10^{-11}  $&$ 23.21124 $&$ 25.02741 $ \\
  $8       $&$ -0.233486 $&$ 16.46697 $&$ -0.000807 $&$ 23.04051 $&$ 25.00957            $&$ -0.000578 $&$ 16.00116 $&$ 1.2 \times 10^{-11}  $&$ 27.81695 $&$ 29.78601 $ \\
  $9       $&$ -0.208700 $&$ 18.41740 $&$ -0.000515 $&$ 28.39540 $&$ 30.50027            $&$ -0.000578 $&$ 18.00116 $&$ 1.1 \times 10^{-11}  $&$ 33.03671 $&$ 35.14158 $ \\
  $10      $&$ -0.188575 $&$ 20.37715 $&$ -0.000344 $&$ 34.35062 $&$ 36.57758            $&$ -0.000579 $&$ 20.00116 $&$ 1.0 \times 10^{-11}  $&$ 38.87051 $&$ 41.09748 $ \\

 \end{tabular}
\end{ruledtabular}
 \end{table*}
 \endgroup

\begingroup
\squeezetable
\begin{table*}
\caption{\label{tab:Table7} Image positions  on the same side as the source due to lensing by the Galactic MDO 
            modeled as a strongly naked singularity ($\nu =0.02$), and their respective bending angles, magnifications, time delays
          and differential time delays.
       $({\rm a})$ The same as (${\rm a}$) of Table II.
       $({\rm b})$ The same as (${\rm b}$)  of Table V.
       $({\rm c})$ The same as  (${\rm a}$) of Table I. }
\begin{ruledtabular}
  \begin{tabular}{l|ccccc|cccc}
\multicolumn{1}{c|}{$\beta$}&
\multicolumn{5}{c|}{Inner image on the same side as the source}&
\multicolumn{4}{c}{Outer image on the same side as the source}\\
&    $\theta_{is}$  &   $\hat{\alpha}_{is}$ &    $\mu_{is}$   &   $\tau_{is}$  & $\Delta\tau_{is}$  &    $\theta_{os}$  &  $\hat{\alpha}_{os}$ &  $\mu_{os}$ & $\tau_{os}$    \\
\hline
  $0       $&$ 0.002297  $&$  0.004593 $&$  \times            $&$ -8.271824 $&$ 5.808667 $&$ 1.387029 $&$ 2.774057 $&$ \times $&$ -14.08049 $\\
 $10^{-6} $&$ 0.002297  $&$  0.004591 $&$ -0.006273           $&$ -8.271824 $&$ 5.808667 $&$ 1.387029 $&$ 2.774056 $&$ 694085.9 $&$ -14.08049 $\\
 $10^{-5} $&$ 0.002297  $&$  0.004573 $&$ -0.000627           $&$ -8.271824 $&$ 5.808675 $&$ 1.387034 $&$ 2.774047 $&$ 69409.04 $&$ -14.08050 $\\
 $10^{-4} $&$ 0.002297  $&$  0.004393 $&$ -0.000063           $&$ -8.271824 $&$ 5.808752 $&$ 1.387079 $&$ 2.773958 $&$ 6941.354 $&$ -14.08058 $\\
 $10^{-3} $&$ 0.002297  $&$  0.002593 $&$ -0.000006           $&$ -8.271825 $&$ 5.809517 $&$ 1.387529 $&$ 2.773058 $&$ 694.5857 $&$ -14.08134 $\\
 $10^{-2} $&$ 0.002297  $&$ -0.015407 $&$ -6.3 \times 10^{-7} $&$ -8.271807 $&$ 5.817185 $&$ 1.392042 $&$ 2.764084 $&$ 69.91009 $&$ -14.08899 $\\
 $10^{-1} $&$ 0.002296  $&$ -0.195407 $&$ -6.3 \times 10^{-8} $&$ -8.268894 $&$ 5.895250 $&$ 1.437970 $&$ 2.675940 $&$ 7.454558 $&$ -14.16414 $\\
 $1       $&$ 0.002294  $&$ -1.995412 $&$ -6.2 \times 10^{-9} $&$ -7.966209 $&$ 6.830681 $&$ 1.974717 $&$ 1.949434 $&$ 1.322625 $&$ -14.79689 $\\
 $2       $&$ 0.002291  $&$ -3.995418 $&$ -3.1 \times 10^{-9} $&$ -7.046546 $&$ 8.261433 $&$ 2.710378 $&$ 1.420757 $&$ 1.073949 $&$ -15.30798 $\\
 $3       $&$ 0.002289  $&$ -5.995423 $&$ -2.1 \times 10^{-9} $&$ -5.512834 $&$ 10.17630 $&$ 3.543471 $&$ 1.086943 $&$ 1.024177 $&$ -15.68914 $\\
 $4       $&$ 0.002286  $&$ -7.995428 $&$ -1.5 \times 10^{-9} $&$ -3.365073 $&$ 12.62205 $&$ 4.434343 $&$ 0.868685 $&$ 1.009733 $&$ -15.98712 $\\
 $5       $&$ 0.002283  $&$ -9.995434 $&$ -1.2 \times 10^{-9} $&$ -0.603263 $&$ 15.62622 $&$ 5.359405 $&$ 0.718809 $&$ 1.004544 $&$ -16.22948 $\\
 $6       $&$ 0.002280  $&$ -11.99544 $&$ -1.0 \times 10^{-9} $&$  2.772596 $&$ 19.20537 $&$ 6.305498 $&$ 0.610996 $&$ 1.002370 $&$ -16.43277 $\\
 $7       $&$ 0.002278  $&$ -13.99544 $&$ -8.6 \times 10^{-10}$&$  6.762504 $&$ 23.36992 $&$ 7.265157 $&$ 0.530315 $&$ 1.001345 $&$ -16.60742 $\\
 $8       $&$ 0.002275  $&$ -15.99545 $&$ -7.5 \times 10^{-10}$&$  11.36646 $&$ 28.12674 $&$ 8.233968 $&$ 0.467935 $&$ 1.000816 $&$ -16.76028 $\\
 $9       $&$ 0.002273  $&$ -17.99545 $&$ -6.6 \times 10^{-10}$&$  16.58447 $&$ 33.48053 $&$ 9.209197 $&$ 0.418394 $&$ 1.000522 $&$ -16.89606 $\\
 $10      $&$ 0.002270  $&$ -19.99546 $&$ -5.9 \times 10^{-10}$&$  22.41652 $&$ 39.43466 $&$ 10.18908 $&$ 0.378166 $&$ 1.000349 $&$ -17.01814$\\
 \end{tabular}
\end{ruledtabular}
 \end{table*}
 \endgroup
\begingroup
\squeezetable
\begin{table*}
\caption{\label{tab:Table8} Image positions  on the opposite  side of the source due to lensing by the Galactic MDO 
            modeled as a strongly naked singularity ($\nu =0.02$), and their respective bending angles, magnifications, time delays 
           and differential time delays.
       $({\rm a})$ The same as (${\rm a}$) of Table II.
       $({\rm b})$ The same as (${\rm b}$)  of Table VI.
       $({\rm c})$ The same as  (${\rm a}$) of Table I.
   }
\begin{ruledtabular}
  \begin{tabular}{l|ccccc|ccccc}
\multicolumn{1}{c|}{$\beta$}&
\multicolumn{5}{c|}{Outer image on the opposite side from the source}&
\multicolumn{5}{c}{Inner image on the opposite side from the source}\\
&    $\theta_{oo}$  & $\hat{\alpha}_{oo}$ & $\mu_{oo}$ &   $\tau_{oo}$ & $\Delta\tau_{oo}$    &    $\theta_{io}$  &  $\hat{\alpha}_{io}$ & $\mu_{io}$ & $\tau_{io}$ & $\Delta\tau_{io}$  \\
\hline
 $0       $&$ -1.387029 $&$ 2.774057 $&$   \times  $&$ -14.08049 $&$ 0          $&$ -0.002297 $&$ 0.004593 $&$ \times             $&$ -8.271824 $&$ 5.808667 $ \\
 $10^{-6} $&$ -1.387028 $&$ 2.774058 $&$ -694084.9 $&$ -14.08049 $&$ 0.000002   $&$ -0.002297 $&$ 0.004595 $&$ 0.006273           $&$ -8.271824 $&$ 5.808667 $ \\
 $10^{-5} $&$ -1.387024 $&$ 2.774067 $&$ -69408.04 $&$ -14.08048 $&$ 0.000017   $&$ -0.002297 $&$ 0.004613 $&$ 0.000627           $&$ -8.271824 $&$ 5.808675 $ \\
 $10^{-4} $&$ -1.386979 $&$ 2.774157 $&$ -6940.354 $&$ -14.08041 $&$ 0.000170   $&$ -0.002297 $&$ 0.004793 $&$ 0.000063           $&$ -8.271824 $&$ 5.808752 $ \\
 $10^{-3} $&$ -1.386528 $&$ 2.775057 $&$ -693.5853 $&$ -14.07964 $&$ 0.001703   $&$ -0.002297 $&$ 0.006593 $&$ 0.000006           $&$ -8.271822 $&$ 5.809520 $ \\
 $10^{-2} $&$ -1.382034 $&$ 2.784067 $&$ -68.90968 $&$ -14.07196 $&$ 0.017034   $&$ -0.002297 $&$ 0.024593 $&$ 6.3 \times 10^{-7} $&$ -8.271779 $&$ 5.817213 $ \\
 $10^{-1} $&$ -1.337888 $&$ 2.875776 $&$ -6.454149 $&$ -13.99377 $&$ 0.170377   $&$ -0.002297 $&$ 0.204594 $&$ 6.3 \times 10^{-8} $&$ -8.268612 $&$ 5.895533 $ \\
 $1       $&$ -0.973942 $&$ 3.947885 $&$ -0.322284 $&$ -13.05731 $&$ 1.739578   $&$ -0.002299 $&$ 2.004599 $&$ 6.3 \times 10^{-9} $&$ -7.963389 $&$ 6.833501 $ \\
 $2       $&$ -0.709041 $&$ 5.418082 $&$ -0.073730 $&$ -11.62581 $&$ 3.682166   $&$ -0.002302 $&$ 4.004604 $&$ 3.2 \times 10^{-9} $&$ -7.040905 $&$ 8.267074 $ \\
 $3       $&$ -0.541791 $&$ 7.083582 $&$ -0.024049 $&$ -9.710471 $&$ 5.978668   $&$ -0.002305 $&$ 6.004610 $&$ 2.1 \times 10^{-9} $&$ -5.504373 $&$ 10.18477 $ \\
 $4       $&$ -0.432460 $&$ 8.864920 $&$ -0.009657 $&$ -7.264419 $&$ 8.722705   $&$ -0.002308 $&$ 8.004615 $&$ 1.6 \times 10^{-9} $&$ -3.353791 $&$ 12.63333 $ \\
 $5       $&$ -0.357399 $&$ 10.71480 $&$ -0.004497 $&$ -4.260046 $&$ 11.96944   $&$ -0.002310 $&$ 10.00462 $&$ 1.3 \times 10^{-9} $&$ -0.589160 $&$ 15.64032 $ \\
 $6       $&$ -0.303414 $&$ 12.60683 $&$ -0.002339 $&$ -0.680750 $&$ 15.75203   $&$ -0.002313 $&$ 12.00463 $&$ 1.1 \times 10^{-9} $&$  2.789520 $&$ 19.22229 $ \\
 $7       $&$ -0.263020 $&$ 14.52604 $&$ -0.001324 $&$  3.483905 $&$ 20.09132   $&$ -0.002316 $&$ 14.00463 $&$ 9.3 \times 10^{-10}$&$  6.782248 $&$ 23.38967 $ \\
 $8       $&$ -0.231793 $&$ 16.46359 $&$ -0.000801 $&$  8.240791 $&$ 25.00107   $&$ -0.002319 $&$ 16.00464 $&$ 8.2 \times 10^{-10}$&$  11.38903 $&$ 28.14930 $ \\
 $9       $&$ -0.206995 $&$ 18.41399 $&$ -0.000511 $&$  13.59463 $&$ 30.49070   $&$ -0.002322 $&$ 18.00464 $&$ 7.4 \times 10^{-10}$&$  16.60985 $&$ 33.50591 $ \\
 $10      $&$ -0.186860 $&$ 20.37372 $&$ -0.000341 $&$  19.54880 $&$ 36.56694   $&$ -0.002325 $&$ 20.00465 $&$ 6.7 \times 10^{-10}$&$  22.44473 $&$ 39.46287 $ \\ 
 \end{tabular}
\end{ruledtabular}
 \end{table*}
 \endgroup
\begingroup
\squeezetable
\begin{table*}
\caption{\label{tab:Table9} Image positions  on the same side as the source due to lensing by the Galactic MDO 
            modeled as the strongly naked singularity ($\nu =0.001$), and their respective bending angles, magnifications, time delays
           and differential time delays.
       $({\rm a})$ The same as (${\rm a}$) of Table II.
       $({\rm b})$ The same as  (${\rm b}$) of Table V.
       $({\rm c})$ The same as  (${\rm a}$) of Table I. 
      }
\begin{ruledtabular}
  \begin{tabular}{l|ccccc|cccc}
\multicolumn{1}{c|}{$\beta$}&
\multicolumn{5}{c|}{ Inner image on the same side as the source}&
\multicolumn{4}{c}{ Outer image on the same side as the source}\\
&    $\theta_{is}$  &  $\hat{\alpha}_{is}$ &    $\mu_{is}$   &   $\tau_{is}$     & $\Delta\tau_{is}$  &    $\theta_{os}$  &  $\hat{\alpha}_{os}$ &    $\mu_{os}$ & $\tau_{os}$    \\
\hline
 $0.75 $&$ \times    $&$ \times   $&$ \times    $&$ \times    $&$ \times   $&$ \times   $&$ \times $&$ \times $&$ \times $\\
 $0.85 $&$ 0.985460 $&$  0.270920 $&$ -1.634301 $&$ -577.4174 $&$ 0.004886 $&$ 1.273212 $&$ 0.846424 $&$ 3.146478 $&$ -577.4223 $\\
 $1    $&$ 0.864695 $&$ -0.270610 $&$ -0.457456 $&$ -577.4165 $&$ 0.049078 $&$ 1.498946 $&$ 0.997892 $&$ 1.854975 $&$ -577.4656 $\\
 $2    $&$ 0.632971 $&$ -2.734059 $&$ -0.038933 $&$ -576.9387 $&$ 0.841932 $&$ 2.488924 $&$ 0.977849 $&$ 1.150355 $&$ -577.7807 $\\
 $3    $&$ 0.542216 $&$ -4.915569 $&$ -0.012362 $&$ -575.7617 $&$ 2.295525 $&$ 3.412890 $&$ 0.825779 $&$ 1.056767 $&$ -578.0572 $\\
 $4    $&$ 0.486179 $&$ -7.027642 $&$ -0.005625 $&$ -573.9272 $&$ 4.363376 $&$ 4.349611 $&$ 0.699223 $&$ 1.026918 $&$ -578.2905 $\\
 $5    $&$ 0.446365 $&$ -9.107269 $&$ -0.003071 $&$ -571.4497 $&$ 7.039872 $&$ 5.300639 $&$ 0.601278 $&$ 1.014622 $&$ -578.4895 $\\
 $6    $&$ 0.415912 $&$ -11.16818 $&$ -0.001874 $&$ -568.3368 $&$ 10.32523 $&$ 6.262637 $&$ 0.525273 $&$ 1.008728 $&$ -578.6620 $\\
 $7    $&$ 0.391513 $&$ -13.21697 $&$ -0.001234 $&$ -564.5931 $&$ 14.22061 $&$ 7.232646 $&$ 0.465292 $&$ 1.005587 $&$ -578.8137 $\\
 $8    $&$ 0.371325 $&$ -15.25735 $&$ -0.000859 $&$ -560.2218 $&$ 18.72713 $&$ 8.208527 $&$ 0.417054 $&$ 1.003774 $&$ -578.9489 $\\
 $9    $&$ 0.354221 $&$ -17.29156 $&$ -0.000623 $&$ -555.2250 $&$ 23.84570 $&$ 9.188781 $&$ 0.377562 $&$ 1.002660 $&$ -579.0707 $\\
 $10   $&$ 0.339464 $&$ -19.32107 $&$ -0.000467 $&$ -549.6044 $&$ 29.57700 $&$ 10.17236 $&$ 0.344712 $&$ 1.001941 $&$ -579.1814 $\\
 $20   $&$ 0.254328 $&$ -39.49134 $&$ -0.000069 $&$ -459.2803 $&$ 120.6604 $&$ 20.09153 $&$ 0.183067 $&$ 1.000238 $&$ -579.9407 $\\
 $25   $&$ 0.231031 $&$ -49.53794 $&$ -0.000037 $&$ -390.9432 $&$ 189.2497 $&$ 25.07404 $&$ 0.148084 $&$ 1.000121 $&$ -580.1930 $\\
 $30   $&$ 0.213366 $&$ -59.57327 $&$ -0.000022 $&$ -307.1928 $&$ 273.2082 $&$ 30.06215 $&$ 0.124291 $&$ 1.000069 $&$ -580.4010 $\\
 $35   $&$ 0.199350 $&$ -69.60130 $&$ -0.000014 $&$ -208.0429 $&$ 372.5350 $&$ 35.05354 $&$ 0.107070 $&$ 1.000043 $&$ -580.5779 $\\
 $40   $&$ 0.187864 $&$ -79.62427 $&$ -0.000010 $&$ -93.50292 $&$ 487.2289 $&$ 40.04702 $&$ 0.094034 $&$ 1.000029 $&$ -580.7318 $\\
 $45   $&$ 0.178218 $&$ -89.64356 $&$ -0.000007 $&$  36.42054 $&$ 617.2886 $&$ 45.04191 $&$ 0.083823 $&$ 1.000020 $&$ -580.8680 $\\
 $50   $&$ 0.169963 $&$ -99.66007 $&$ -0.000005 $&$  181.7226 $&$ 762.7128 $&$ 50.03781 $&$ 0.075611 $&$ 1.000015 $&$ -580.9902 $\\
\end{tabular}
\end{ruledtabular}
\end{table*}
\endgroup
\begingroup
\squeezetable
\begin{table*}
\caption{\label{tab:Table10} Magnification centroid due to lensing by the Schwarzschild black hole ($\nu=1$); and  weakly ($\nu=0.7$), marginally
                strongly ($\nu=0.5$) and strongly ($\nu=0.04, 0.02, 0.001$) naked singularities.
           $({\rm a})$ The same as  (${\rm a}$) of Table I. 
          } 
\begin{ruledtabular}
 \begin{tabular}{l|llllll}
\multicolumn{1}{c|}{ $\beta$}&
\multicolumn{6}{c}{Magnification centroid}\\
&    $\nu=1 \text{(SBH)}$           &         $\nu=0.7 \text{(WNS)}$    &    $\nu=0.5 \text{ (MSNS) }$    &   $\nu=0.04 \text{(SNS)}$    &   $\nu=0.02 \text{(SNS)}$     & $\nu=0.001 \text{(SNS)} $    \\
\hline
 \hline\noalign{\smallskip}
$10^{-6}  $&$ 1.50000000005 \times 10^{-6} $&$ 1.50000000005 \times 10^{-6} $&$ 1.50000000004 \times 10^{-6} $&$ 1.49999993434 \times 10^{-6}   $&$ 1.49999895074 \times 10^{-6}  $&$ \times  $ \\
 $10^{-5} $&$ 0.0000150000000004           $&$ 0.0000150000000003           $&$ 0.0000150000000003           $&$ 0.0000149999993433             $&$ 0.0000149999895073            $&$ \times  $ \\
 $10^{-4} $&$ 0.000149999999875            $&$ 0.000149999999875            $&$ 0.000149999999874            $&$ 0.000149999993304              $&$ 0.000149999894944              $&$ \times  $ \\
 $10^{-3} $&$ 0.00149999987032             $&$ 0.00149999987031             $&$ 0.00149999987031             $&$ 0.00149999980461               $&$ 0.00149999882101               $&$ \times  $ \\
 $10^{-2} $&$ 0.0149998702695              $&$ 0.0149998702694              $&$ 0.0149998702694              $&$ 0.0149998696124                $&$ 0.0149998597771                $&$ \times  $ \\
 $10^{-1} $&$ 0.149870601362               $&$ 0.149870601362               $&$ 0.149870601361               $&$ 0.149870594833                 $&$ 0.149870497099                 $&$ \times  $ \\
 $1       $&$ 1.39699276886                $&$ 1.39699276886                $&$ 1.39699276885                $&$ 1.39699273178                  $&$ 1.39699217730                  $&$ 1.37347550011  $ \\
 $2       $&$ 2.49070719863                $&$ 2.49070719862                $&$ 2.49070719862                $&$ 2.49070717680                  $&$ 2.49070684878                  $&$ 2.42816768265  $ \\
 $3       $&$ 3.44974557083                $&$ 3.44974557083                $&$ 3.44974557082                $&$ 3.44974556168                  $&$ 3.44974541929                  $&$ 3.37969558708  $ \\
 $4       $&$ 4.38823639445                $&$ 4.38823639445                $&$ 4.38823639445                $&$ 4.38823639045                  $&$ 4.38823632168                  $&$ 4.32856338584  $ \\
 $5       $&$ 5.33392461984                $&$ 5.33392461984                $&$ 5.33392461984                $&$ 5.33392461783                  $&$ 5.33392457682                  $&$ 5.28599274987  $ \\
 $6       $&$ 6.29011081174                $&$ 6.29011081174                $&$ 6.29011081174                $&$ 6.29011081057                  $&$ 6.29011078080                  $&$ 6.25179502625  $ \\
 $7       $&$ 7.25521420045                $&$ 7.25521420045                $&$ 7.25521420045                $&$ 7.25521419966                  $&$ 7.25521417489                  $&$ 7.22426206133  $ \\
 $8       $&$ 8.22719524592                $&$ 8.22719524592                $&$ 8.22719524592                $&$ 8.22719524533                  $&$ 8.22719522297                  $&$ 8.20182922137  $ \\
 $9       $&$ 9.20438778298                $&$ 9.20438778298                $&$ 9.20438778298                $&$ 9.20438778249                  $&$ 9.20438776138                  $&$ 9.18329481246  $ \\
 $10      $&$ 10.1855502489                $&$ 10.1855502489                $&$ 10.1855502489                $&$ 10.1855502484                  $&$ 10.1855502280                  $&$ 10.1677715380  $ \\

\end{tabular}
\end{ruledtabular}
\end{table*}
\endgroup
\begingroup
\squeezetable
\begin{table*}
\caption{\label{tab:Table11} Magnification centroid  shift due to lensing by the Schwarzschild black hole ($\nu=1$); and  weakly ($\nu=0.7$), marginally strongly
                           ($\nu=0.5$) and strongly ($\nu= 0.04,0.02, 0.001$) 
                           naked singularities.
       $({\rm a})$ The same as  (${\rm a}$) of Table I.
            }
\begin{ruledtabular}
 \begin{tabular}{l|llllll}
\multicolumn{1}{c|}{ $\beta$}&
\multicolumn{6}{c}{ Magnification centroid shift }\\
&    $\nu=1 \text{(SBH)}$           &         $\nu=0.7 \text{(WNS)}$         &    $\nu=0.5 \text{ (MSNS) }$    &   $\nu=0.04 \text{(SNS)}$    &   $\nu=0.02 \text{(SNS)}$     & $\nu=0.001 \text{(SNS)} $    \\
\hline
\hline\noalign{\smallskip}
 $10^{-6}  $&$   5.00000000050 \times 10^{-7} $&$   5.00000000047 \times 10^{-7}   $&$ 5.00000000040 \times 10^{-7} $&$ 4.99999934340 \times 10^{-7}  $&$ 4.99998950741 \times 10^{-7} $&$ \times              $\\
 $10^{-5}  $&$   5.00000000037 \times 10^{-6} $&$   5.00000000034 \times 10^{-6}   $&$ 5.00000000028 \times 10^{-6} $&$ 4.99999934327 \times 10^{-6}  $&$ 4.99998950728 \times 10^{-6} $&$ \times              $\\
 $10^{-4}  $&$   0.0000499999998753           $&$   0.0000499999998750             $&$   0.0000499999998743         $&$ 0.0000499999933043            $&$   0.0000499998949444         $&$   \times            $\\
 $10^{-3}  $&$   0.000499999870316            $&$   0.000499999870313              $&$   0.000499999870306          $&$ 0.000499999804606             $&$   0.000499998821007          $&$   \times            $\\
 $10^{-2}  $&$   0.00499987026947             $&$   0.00499987026944               $&$   0.00499987026938           $&$ 0.00499986961242              $&$   0.00499985977706           $&$   \times            $\\
 $10^{-1}  $&$   0.0498706013618              $&$   0.0498706013615                $&$   0.0498706013609            $&$ 0.0498705948326               $&$   0.0498704970987            $&$   \times            $\\
 $1        $&$   0.396992768862               $&$   0.396992768859                 $&$   0.396992768855             $&$ 0.396992731780                $&$   0.396992177305             $&$   0.373475500109    $\\
 $2        $&$   0.490707198627               $&$   0.490707198624                 $&$   0.490707198620             $&$ 0.490707176798                $&$   0.490706848782             $&$   0.428167682651    $\\
 $3        $&$   0.449745570828               $&$   0.449745570827                 $&$   0.449745570824             $&$ 0.449745561683                $&$   0.449745419290             $&$   0.379695587076    $\\
 $4        $&$   0.388236394450               $&$   0.388236394449                 $&$   0.388236394447             $&$ 0.388236390447                $&$   0.388236321681             $&$   0.328563385839    $\\
 $5        $&$   0.333924619843               $&$   0.333924619842                 $&$   0.333924619841             $&$ 0.333924617833                $&$   0.333924576816             $&$   0.285992749873    $\\
 $6        $&$   0.290110811742               $&$   0.290110811742                 $&$   0.290110811741             $&$ 0.290110810567                $&$   0.290110780796             $&$   0.251795026246    $\\
 $7        $&$   0.255214200451               $&$   0.255214200451                 $&$   0.255214200450             $&$ 0.255214199661                $&$   0.255214174885             $&$   0.224262061328    $\\
 $8        $&$   0.227195245922               $&$   0.227195245922                 $&$   0.227195245921             $&$ 0.227195245327                $&$   0.227195222966             $&$   0.201829221370    $\\
 $9        $&$   0.204387782975               $&$   0.204387782975                 $&$   0.204387782975             $&$ 0.204387782486                $&$   0.204387761378             $&$   0.183294812456    $\\
 $10       $&$   0.185550248860               $&$   0.185550248859                 $&$   0.185550248859             $&$ 0.185550248432                $&$   0.185550228011             $&$   0.167771538043    $\\

\end{tabular}
\end{ruledtabular}
\end{table*}
\endgroup
\begingroup
\squeezetable
\begin{table*}
\caption{\label{tab:Table12}  Total  magnification due to lensing by the Schwarzschild black hole ($\nu=1$); and  weakly ($\nu=0.7$), marginally 
                              strongly ($\nu=0.5$) and strongly ($\nu=0.04,0.02, 0.001$) naked singularities.
           $({\rm a})$ The same as  (${\rm a}$) of Table I. 
          } 

\begin{ruledtabular}
\begin{tabular}{l|llllll}
\multicolumn{1}{c|}{$\beta$}&
\multicolumn{6}{c}{Total  magnification }\\
&    $\nu=1 \text{(SBH)}$           &         $\nu=0.7 \text{(WNS)}$         &    $\nu=0.5 \text{ (MSNS) }$    &   $\nu=0.04 \text{(SNS)}$  &   $\nu=0.02 \text{(SNS)}$     & $\nu=0.001 \text{(SNS)} $    \\
\hline
\hline\noalign{\smallskip}
 $10^{-6}  $&$ 1.38816932538\times 10^6$&$1.38816932538\times 10^6$&$1.38816932539\times 10^6$&$ 1.38816941645 \times 10^6  $&$ 1.38817077256\times 10^6 $&$\times $\\
 $10^{-5}  $&$ 138816.932541           $&$138816.932541           $&$ 138816.932542          $&$ 138816.941647              $&$ 138817.077259            $&$ \times $\\
 $10^{-4}  $&$ 13881.6932808           $&$ 13881.6932809          $&$ 13881.6932809          $&$ 13881.6941915              $&$ 13881.7077527            $&$ \times $\\
 $10^{-3}  $&$ 1388.16959552           $&$ 1388.16959552          $&$ 1388.16959553          $&$ 1388.16968659              $&$ 1388.17104270            $&$ \times $\\
 $10^{-2}  $&$ 138.819633923           $&$ 138.819633923          $&$ 138.819633924          $&$ 138.819643029              $&$ 138.819778637            $&$ \times $\\
 $10^{-1}  $&$ 13.9086926599           $&$ 13.9086926600          $&$ 13.9086926601          $&$ 13.9086935678              $&$ 13.9087070859            $&$ \times $\\
 $1        $&$ 1.64490824157           $&$ 1.64490824157          $&$ 1.64490824158          $&$ 1.64490830940              $&$ 1.64490931584            $&$ 2.31243076497 $\\
 $2        $&$ 1.14767800038           $&$ 1.14767800038          $&$ 1.14767800038          $&$ 1.14767801708              $&$ 1.14767826338            $&$ 1.18928813001 $\\
 $3        $&$ 1.04822623968           $&$ 1.04822623968          $&$ 1.04822623968          $&$ 1.04822624452              $&$ 1.04822631629            $&$ 1.06912991460 $\\
 $4        $&$ 1.01939069971           $&$ 1.01939069971          $&$ 1.01939069971          $&$ 1.01939070136              $&$ 1.01939072680            $&$ 1.03254298220 $\\
 $5        $&$ 1.00904161659           $&$ 1.00904161659          $&$ 1.00904161659          $&$ 1.00904161725              $&$ 1.00904162855            $&$ 1.01769251767 $\\
 $6        $&$ 1.00470884575           $&$ 1.00470884575          $&$ 1.00470884575          $&$ 1.00470884606              $&$ 1.00470885230            $&$ 1.01060207986 $\\
 $7        $&$ 1.00266919134           $&$ 1.00266919134          $&$ 1.00266919134          $&$ 1.00266919151              $&$ 1.00266919562            $&$ 1.00682076296 $\\
 $8        $&$ 1.00161696592           $&$ 1.00161696592          $&$ 1.00161696592          $&$ 1.00161696603              $&$ 1.00161696910            $&$ 1.00463247208 $\\
 $9        $&$ 1.00103306445           $&$ 1.00103306445          $&$ 1.00103306445          $&$ 1.00103306453              $&$ 1.00103306701            $&$ 1.00328346259 $\\
 $10       $&$ 1.00068928992           $&$ 1.00068928992          $&$ 1.00068928992          $&$ 1.00068928998              $&$ 1.00068929209            $&$ 1.00240885784 $\\
\end{tabular}
\end{ruledtabular}
\end{table*}
\endgroup    

\begin{figure*}
  \renewcommand{\topfraction}{2} 
 \centerline{ \epsfxsize 14cm
   \epsfbox{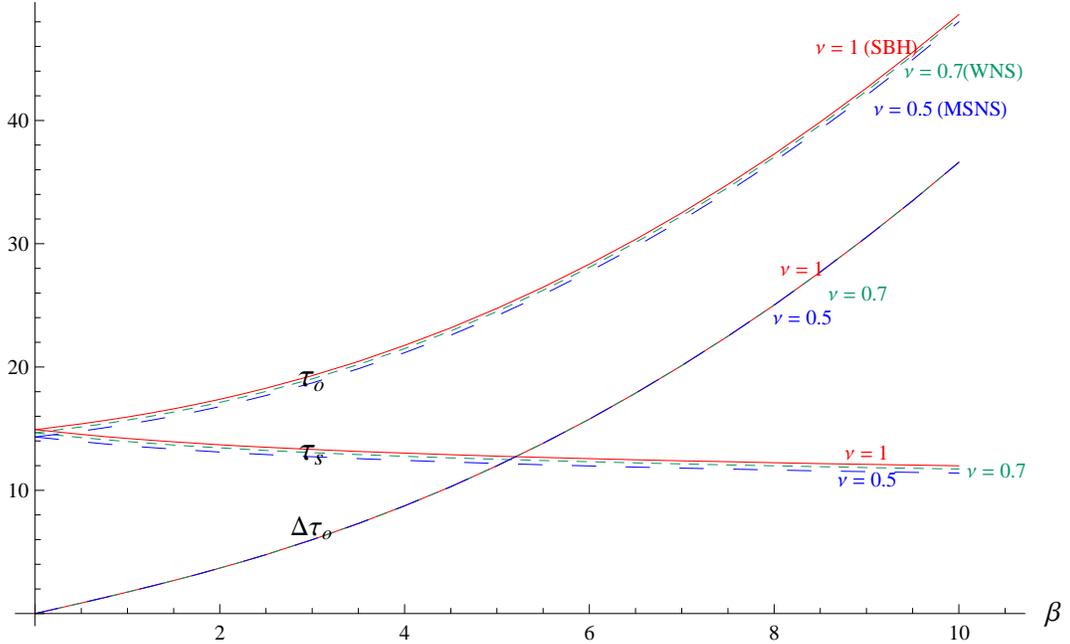}}
 \caption[ ]
{ The {\em time delays} of the images on the opposite side from the source $\tau_o$, the same side as the source
  $\tau_s$, and the {\em differential time delay} of the images on the opposite side from the source $\Delta\tau_o$
   are plotted against the angular source position $\beta$ for $\nu=1$ (SBH),  $\nu=0.7$ (WNS), and $\nu=0.5$ (MSNS).
 $M/D_d \approx 2.26 \times  10^{-11} $ and  $D_{ds}/D_s=1/2$.  
     The time delays as well as differential time delays are expressed in minutes whereas the angular source position is
     given in  arcseconds.
}
\label{fig1}
\end{figure*}
\begin{figure*}
\centerline{ \epsfxsize 12cm
   \epsfbox{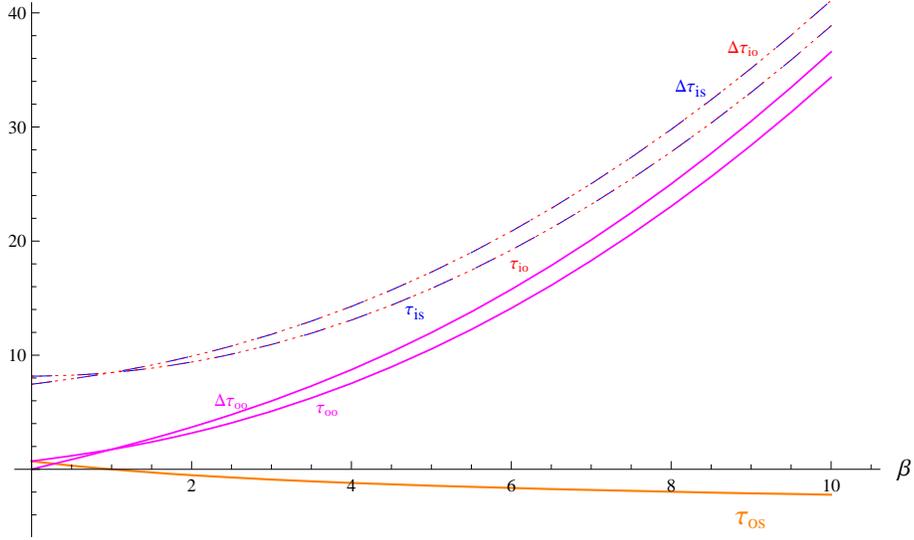}}
 \caption[ ]
{The {\em time delays} of the outer image on the same side as the source $\tau_{os}$, inner image on the same side as the
source $\tau_{is}$, inner image on the opposite side from the source $\tau_{io}$, and outer image on the opposite side from the 
source $\tau_{oo}$ are plotted against the angular source position $\beta$ for $\nu=0.04$ (SNS). Also,  the {\em differential time delays} of the
inner images on the same side as the source $\Delta\tau_{is}$, inner images on the opposite side from the source $\Delta\tau_{io}$,
and outer  images on the opposite side from the source $\Delta\tau_{oo}$ are plotted against $\beta$ for the same value of $\nu$.
$D_{ds}/D_s=1/2$ and  $M/D_d \approx 2.26 \times  10^{-11}$. The time delays as well as differential time delays are 
in minutes whereas the angular source position is expressed in arcseconds.
 }
\label{fig2}
\end{figure*}
 \begin{figure*}
\centerline{ \epsfxsize 12cm
   \epsfbox{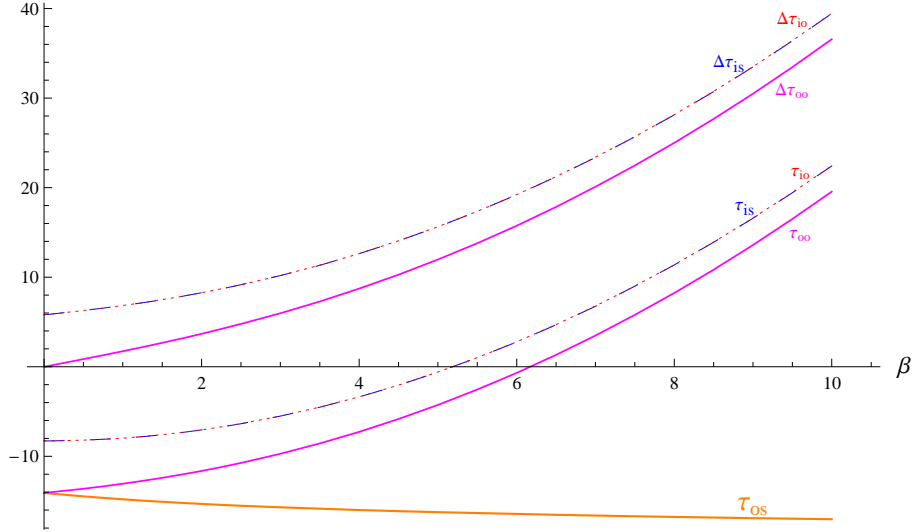}}
 \caption[ ]
{The {\em time delays} of the direct image $\tau_{os}$, inner image on the same side as the
source $\tau_{is}$, inner image on the opposite side from the source $\tau_{io}$, and outer image on the opposite side from the 
source $\tau_{oo}$ are plotted against the angular source position $\beta$ for $\nu=0.02$ (SNS). The {\em differential time delays} of the
inner images on the same side as the source $\Delta\tau_{is}$, inner images on the opposite side from the source $\Delta\tau_{io}$,
and outer  images on the opposite side from the source $\Delta\tau_{oo}$ are plotted against $\beta$ for the same value of $\nu$.
The angular source position is expressed in arcseconds whereas the time delays as well as differential time delays are shown
in minutes. $M/D_d \approx 2.26 \times  10^{-11}$ and $D_{ds}/D_s=1/2$. 
}
\label{fig3}
\end{figure*}

 \begin{figure*}
\centerline{ \epsfxsize 16cm
   \epsfbox{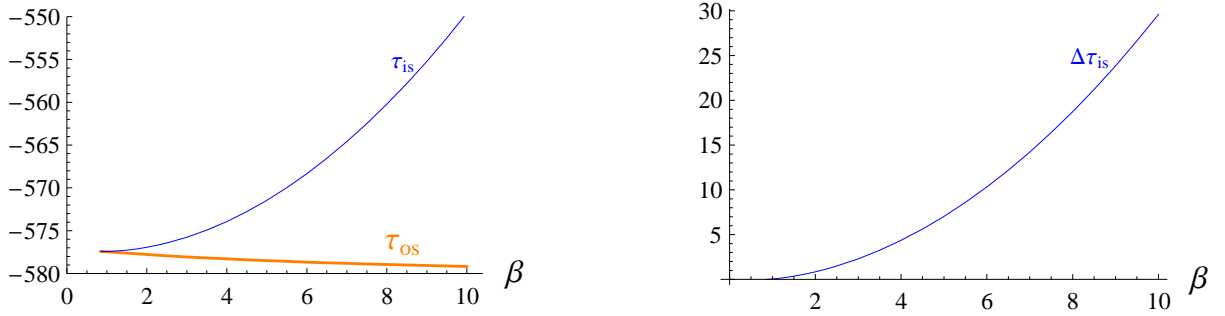}}
 \caption[ ]
{The {\em time delays} of the outer images on the same side as the source $\tau_{os}$ and  inner images on the same side as the
source $\tau_{is}$ are plotted against the angular source position $\beta$  for $\nu=0.001$ (SNS); see the left hand
side figure. The {\em differential time delay} of the inner image on the same side as the source $\tau_{is}$ 
is  plotted against  $\beta$  (right hand side figure).  The time delays as well as differential time delays
and  the angular source position are respectively given in minutes and arcseconds. $D_{ds}/D_s=1/2$ and  $M/D_d \approx 2.26 \times  10^{-11}$.
}
\label{fig4}
\end{figure*}

 \begin{figure*}
 
\centerline{ \epsfxsize 18cm
   \epsfbox{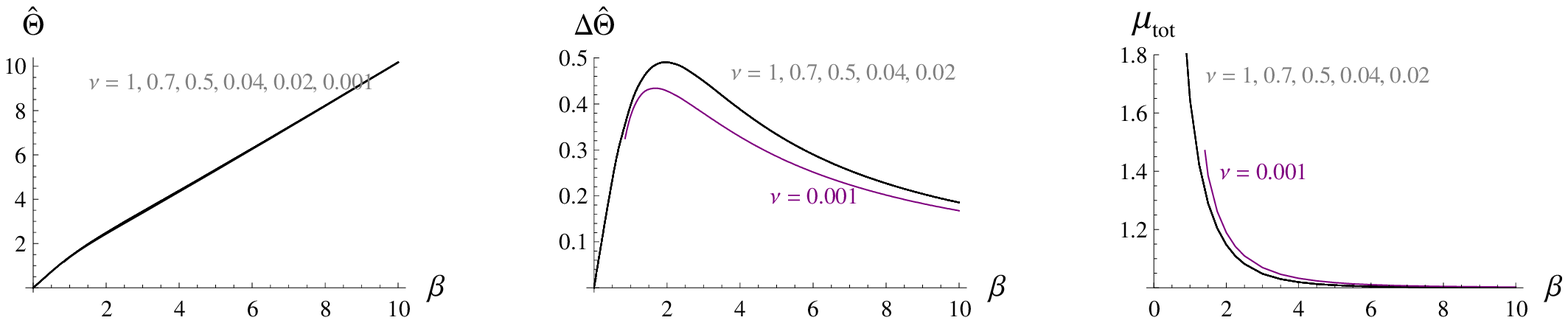}}
 \caption[ ]
{The {\em magnification centroid} $\hat{\Theta}$,  {\em magnification centroid shift} $\Delta\hat{\Theta}$, and  the {\em total
 magnification} $\mu_{tot}$  are plotted against the angular source position $\beta$  for 
$\nu=1$ (SBH), $\nu=0.7$ (WNS), $\nu=0.5$ (MSNS), and $\nu= 0.04, 0.02, 0.001$ (SNS).
$D_{ds}/D_s=1/2$ and  $M/D_d \approx 2.26 \times  10^{-11}$. $\hat{\Theta}$, $\Delta\hat{\Theta}$, and $\beta$ are
expressed in arcseconds.
}
\label{fig5}
\end{figure*}

As shown in Table I, there is only one Einstein ring and no RCC for the case of SBH ($\nu=1$), WNS ($\nu=0.9, 0.8, 0.7, 
0.6$), and MSNS ($\nu=0.5$).  The angular positions of the Einstein rings decrease very slowly with a 
decrease in the value of $\nu$ (equivalently, an increase in  the value of $(q/M)^2$). In the case of SNS lensing
($0\leq \nu < 1/2$), there are two situations: There is always one RCC; however, there can be two (for example, for $\nu 
= 0.4, 0.3$, etc.) or none Einstein rings (for example, for $\nu = 0.001$ and any lower but nonnegative values of $\nu$).
For the cases of double Einstein rings, the angular radii of outer and inner rings, respectively, decrease and increase
with a decrease in the value of $\nu$.  For  a  detailed  analysis of these RCCs and TCCs, see \cite{VE02}.

We now compute image positions, the corresponding deflection angles of light, magnifications, time delays, and differential
time delays for several values of the angular source position for different values of $\nu=1$ (SBH), $\nu=0.7$ (WNS), 
$\nu=0.5$ (MSNS), and $\nu=0.04, 0.02, 0.001$ (SNS). Though we give only a few data in the tables, we have computed and
used  many more data points for the figures.

The gravitational lensing effects due to the SBH, WNS, and MSNS are qualitatively similar, though they differ quantitatively by
small values (see Tables {\rm II} through {\rm IV} and Figure 1).  For each case, there are two images, one on each side of the 
optical axis. As the source moves away from the optical axis,
the image on the same side  as the source (i.e., the direct image) moves away from the  axis, whereas the image on the opposite 
side of the source moves 
toward the axis. The absolute magnification of both images decreases. 
For a given value of $\nu$, with an increase in the angular position of the source, the time delays of the images
on the same side as the source decrease, whereas the time delays and differential time delays of the image on the opposite side from the
source increase. The rate of decrease in time delay of the image on the same
side as the source is much slower than the rate of increase in time delay of the image on the opposite side from the source.
For any given value of the angular source position, the time delays of both images and the differential time delay of the image on the
opposite side from the source decrease with a decrease in the  value of $\nu$ (equivalently, increase in $(q/M)^2$).

Gravitational lensing by a SNS is qualitatively {\em very} different from lensing by SBH, WNS, or MSNS (see Tables
${\rm V}$ through ${\rm IX}$ and Figures 2 through 4). For $\nu=0.04$ and  $0.02$, when the lens components (the source, lens 
and observer) are
perfectly (or nearly) aligned, there are two concentric Einstein rings (the inner ring  much fainter than the outer one).
As the alignment is broken, the  two Einstein rings break into  four images, two images on each side of the optical axis.
The time delay of  direct  image decreases with an increase in $\beta$.  For $\nu=0.04$, the time delay of the direct image is 
positive for small $\beta$ and negative for large $\beta$; however, the time delays for other images are  positive for all values
of $\beta$.
On the other hand, for $\nu=0.02$, the time delay of the direct image is always negative; however, the other three images have negative
time delays for small $\beta$ and positive for large $\beta$, passing through the zero time delay point.

 For a given value of $\beta$, the time
delays of images are in the following decreasing order: the inner image on the opposite side from the source, inner image
on the same side as the source, outer image on the opposite side from the source, and the direct image. However,  for any
given value of $\beta$, the absolute 
magnifications of images are not in the exact reverse sequence; they are rather in the following decreasing order: 
The direct image, outer image on the opposite side from the source, inner image on the  opposite side from the source, and
inner image on the same side as the source.  The differential time delays are 
always positive. It is worth emphasizing that the negative and positive time delays are  respectively not necessarily
due to negative and positive bending angles. A light ray with a positive deflection angle may give rise to positive or 
negative time delays, and the same is true for a light ray with a negative deflection angle. If $\beta$ increases, the
angular separation between images on the same side as the source increases (the inner and outer images move respectively
toward and away from the optical  axis). However, the
angular separation between images on the opposite side from the source decreases (the outer and inner images move respectively
toward and away from the optical axis) and for some value of $\beta$ these two images 
coalesce to form one  highly magnified image; for example, for $\nu=0.02$, the angular positions of the source and image are $\beta  \approx  210.2934$ \ 
arcseconds (RC) and $\theta \approx  - 0.004591$ arcsecond (RCC), respectively. The opposite signs on these two values
 show that the RCC and RC 
are on the opposite sides of optical axis (see Table ${\rm I}$). For further increase in $\beta$,  there are no images
on the opposite side from the source.

For $\nu=0.001$,  see Table ${\rm IX}$ and Figure 4. If the lens components are  aligned, there are no Einstein rings. 
Also, for small values of $\beta$, there
are no images on either side of the optical axis. For $\beta \approx 0.808397$ \  arcsecond (RC), there is a highly magnified 
image at the angular position $\theta \approx 1.115015$  arcseconds (RCC). The signs on RC and RCC being the same implies that this
image appear on  the same side as the source. As $\beta$ further increases, the image splits into two and the inner and outer 
(direct)
images move  toward and away from the optical axis, respectively; the magnification of inner image decreases much faster than
the outer one. The time delay of the direct image for this case is always negative and decreases slowly with an increase in $\beta$.
The time delay of the inner image, however, is negative for small $\beta$ and positive for large $\beta$, passing through the
zero time delay for a certain value of $\beta$. The increase rate of time delay of the inner image is much higher than the
decrease rate of time delay of the direct image. 
For the inner image, the deflection angles are  positive and negative respectively  for  small and large values
of $\beta$; however, for the direct image it is always positive.
As for other cases already discussed, the differential time delay of the inner image 
is always positive.

We now compute the magnification centroid, magnification centroid shift, and the total  magnification for $\nu=1$(SBH),
 $\nu=0.7$ (WNS), $\nu=0.5$ (MSNS),  and $\nu=0.04, 0.02, 0.001$ (SNS) for several values of $\beta$ (see Tables ${\rm X}$ through
${\rm XII}$). We then plot these quantities against $\beta$ (see Fig. 5). For a fixed value of $\nu$, the magnification centroid increases with an
increase in $\beta$. For small values of $\beta$, the graph is bulged up and then tends to become straight as $\beta$  increases.
For a fixed value of $\beta$, the magnification centroid decreases with a decrease in $\nu$ (i.e., increase in $(q/M)^2$);
the decrease is however too small  for  these five graphs to appear resolved even if these were plotted on an entire page.
As $\beta$ increases, the  magnification centroid shift first increases, reaches a maximum value, and then starts decreasing 
to the limiting value zero. For a given value of $\beta$, the  magnification centroid shift decreases with a decrease in the
 value of $\nu$. For a given value of $\nu$, the total  magnification is, as expected, very high for small $\beta$ and it decreases
to the limiting value one,  as  $\beta$ increases. For any given value of $\beta$, however, the total magnification 
increases with a decrease in the value of $\nu$. Thus, the presence of scalar charge helps increase the total magnification.
This would provide  a modest increase in the likelihood of observing lensing by Sgr $A^{*}$ (see \cite{CKN07} and references
therein).

All the images produced by the Galactic MDO (modeled as a Schwarzschild black hole and  naked singularity lenses) may be 
resolved from each other by observational facilities available in the near future. Therefore, the results of magnification 
centroid and total magnification may not be needed. However, our studies help us understand the role of scalar field on the 
magnification centroid and total magnification, which could be useful while studying gravitational lensing by exotic 
 dark objects having  rather small value of $M/D_d$.


\section{\label{sec:Discussion&Conclusion} Discussion and Conclusion}

The naked singularities are classified in three categories: {\em WNS}, {\em MSNS}, and {SNS}. We modeled the Galactic MDO 
as the SBH, and Janis-Newman-Winicour WNS, MSNS and SNS lenses, and studied point source gravitational lensing by them. 
We found that the  gravitational lensing
effects due to the SBH, WNS, and MSNS are qualitatively similar (but these differ slightly quantitatively) to  each other;
however, they differ qualitatively from SNS lensing. Therefore, it will be easier to observationally differentiate a SNS
(compared to  a WNS or a MSNS) from a SBH.

{\em SBH, WNS and MSNS lensing}. 
These do not give rise to any radial caustics; however, they do produce one Einstein ring
when the lens components are aligned (i.e. $\beta=0$).  When $\beta$ increases, the Einstein ring splits into two images, 
one on each side of the optical axis. The time delays for both images are positive for all values of $\beta$. For a given
value of $\beta$, a decrease in $\nu$, i.e., an increase in $(q/M)^2$, decreases the absolute angular image positions, 
time delays, magnification centroid, and magnification centroid shift; however, it increases the magnitude of  magnifications
of images and therefore also the total  magnification. The differences are however very small. 
The deflection angle $\hat{\alpha}$ becomes unboundedly large as the impact parameter $J \rightarrow J_{ps}$ for the SBH and WNS,
and $ J \rightarrow 0$ for the MSNS \cite{VE02}.

{\em SNS lensing}. 
There are  two types of lensing in this category. In the first, for example for the case of  $\nu = 0.02$,  there are double 
concentric Einstein rings (when $\beta = 0$) and one  radial critical curve (when $\beta \approx 210.2934$   arcseconds).
As the angular position of the source  increases from the alignment position of the lens components (i.e., $\beta = 0$), 
the two Einstein rings ``break'' into four images,  giving two images on each side of the optical axis.  The separation
 between images on the same and opposite side from the source, respectively, increases and decreases as $\beta$ increases, 
and eventually the two images on the opposite side from the source coalesce to form a single highly magnified image. 
For any  further increase in $\beta$, there are
only two images on the same side as the source. 
The second category of SNS lensing, for example for  $\nu=0.001$, there is one RC; however, there is no
Einstein ring when $\beta=0$. Moreover, there is no image for small values of $\beta$. As $\beta$ increases,  a highly
magnified image (RCC) first appears  on the same side as the source. A further increase in $\beta$ splits this into two images and 
the separation between them keeps increasing  (both images remaining on  the same side as the source). 
The time delay of  images of SNS lensing may be positive, zero, or negative depending on the values of $\nu$ and $\beta$. However,  
the time delay of a direct image is negative
for any SNS lensing if $\beta$ is large.
As shown in \cite{VE02}, the deflection angle $\hat{\alpha}$ approaches $-\pi$ as the
impact parameter approaches its minimum value zero. Therefore, if a light ray is sent radially toward a SNS, it will 
``bounce''; however, it remains to be computed whether or not the ``reflected'' light has enough magnification to be observed
by present instruments or by those likely to be available in near future. This may serve as a crucial test for the existence of SNS.

The naked singularity lensing gives rise to images of smaller time delay and stronger magnification than the black hole lensing.
Therefore, if naked singularities indeed exist in nature, then these will serve as {\em better cosmic telescopes} and  will 
help probe the universe more efficiently. 
The results obtained in this paper also helps understand the effects of scalar field on gravitational lensing, which could have valuable 
implications for research in  cosmology.

The Janis-Newman-Winicour metric also describes the exterior gravitational
field of a scalar star. Therefore, results obtained in this paper for naked singularities are also applicable to
scalar stars. The scalar star however must be compact enough for the images not to be obstructed.

The  metric we considered in this paper may or may not be physically realistic. However,  gravitational 
lensing studies with this metric serves as a  stepping stone to understand the distinctive lensing features of black holes and naked 
singularities. 
It would be indeed of great astrophysical significance to obtain distinguishing  {\em qualitative} lensing 
characteristics of Kerr black holes and naked singularities, so that the  weak cosmic censorship hypothesis could be tested 
observationally  without any ambiguity.

\acknowledgments
Thanks are due to J.~M.~Aguirregabiria of the University of the Basque Country (Spain)  for some helpful corresponsence related to  Mathematica.


\newpage

\begin{references}
\bibitem{Pen98}
           R.~Penrose, in   {\em Black Holes and Relativistic Stars},
               edited by R. M. Wald (The University of Chicago Press,
               Chicago, 1998), p.~103.

\bibitem{Wal97}
           R.~M.~ Wald, ``Gravitational  Collapse and Cosmic Censorship,''
           gr-qc/9710068.

\bibitem{Vir99}
          K.~S.~Virbhadra, Phys. Rev. D {\bf 60}, 104041 (1999).

\bibitem{VNC98}
     K.~S.~Virbhadra, D.~Narasimha, and S.~M.~Chitre, Astron.  Astrophys.
               {\bf 337}, 1 (1998).

\bibitem{VE00}
      K.~S.~Virbhadra and G.~F.~R.~Ellis,   Phys. Rev. D {\bf 62}, 084003
            (2000). The light deflection in the vicinity of a photon sphere of a Schwarzschild black hole was also 
             previously studied by some researchers; for example,   
      C.~Darwin, Proc. R. Soc. London A {\bf 249}, 180 (1958); 
                              {\it ibid} {\bf 263}, 39 (1961);
      H.~C.~Ohanian, Am. J. Phys. {\bf 55}, 428 (1987);
      R.~J.~Nemiroff, {\it ibid}, {\bf 61}, 619 (1993).

\bibitem{CVE01}
     C-M.~Claudel, K.~S.~Virbhadra, and G.~F.~R.~Ellis, J. Math. Phys.
              {\bf 42}, 818 (2001).


\bibitem{VE02}
      K.~S.~Virbhadra and G.~F.~R.~Ellis,   Phys. Rev. D {\bf 65}, 103004
            (2002).



\bibitem{Per} 
       V.~Perlick, Living Rev.\ Rel.\  {\bf 7}, 9 (2004);
                   Commun.\ Math.\ Phys.\  {\bf 220}, 403 (2001);
                   Phys.\ Rev.\  D {\bf 69}, 064017 (2004);
           ``Theoretical gravitational lensing. Beyond the weak-field small-angle
             approximation,'' arXiv:0708.0178 [gr-qc];
        W.~Hasse and V.~Perlick,
                   Gen.\ Rel.\ Grav.\  {\bf 34}, 415 (2002).

\bibitem{Zaketal} 
       A.~F.~Zakharov and Yu.~V.~Baryshev,
                    Class.\ Quant.\ Grav.\  {\bf 19}, 1361 (2002);
                    Int.\ J.\ Mod.\ Phys.\  D {\bf 11}, 1067 (2002);
      A.~F.~Zakharov, F.~De Paolis, G.~Ingrosso and A.~A.~Nucita,   
                    New Astron.\  {\bf 10}, 479 (2005).
     
  
 \bibitem{Amoreetal} 
      
  P.~Amore and S.~Arceo Diaz,  
              Phys.\ Rev.\  D {\bf 73}, 083004 (2006);  
  P.~Amore, S.~Arceo and F.~M.~Fernandez,
                 {\it ibid}   {\bf 74}, 083004 (2006);
  P.~Amore, M.~Cervantes, A.~De Pace and F.~M.~Fernandez,
               {\it ibid}  {\bf 75}, 083005 (2007).


 \bibitem{Bozzaetal} 
    V.~Bozza, S.~Capozziello, G.~Iovane and G.~Scarpetta,
              Gen.\ Rel.\ Grav.\  {\bf 33}, 1535 (2001);
    V.~Bozza,
         Phys.\ Rev.\  D  {\bf 66}, 103001 (2002);
              {\it ibid}  {\bf 67}, 103006 (2003);             
     V.~Bozza, F.~De Luca and G.~Scarpetta,  
               {\it ibid} {\bf 74}, 063001 (2006);
     V.~Bozza and L.~Mancini,
         Astrophys.\ J.\  {\bf 627}, 790 (2005).


 \bibitem{Eioraetal} 
   
     E.~F.~Eiroa, G.~E.~Romero and D.~F.~Torres,
                  Phys.\ Rev.\  D {\bf 66}, 024010 (2002);
     E.~F.~Eiroa and D.~F.~Torres,
                  {\it ibid}     {\bf 69}, 063004 (2004);
     E.~F.~Eiroa,  Braz.\ J.\ Phys.\  {\bf 35}, 1113 (2005);  
                 Phys.\ Rev.\  D {\bf 73}, 043002 (2006).
   
\bibitem{KNPF} 
    T.~P.~Kling, E.~T.~Newman and A.~Perez,
                 Phys.\ Rev.\  D {\bf 61}, 104007 (2000);
                 {\it ibid}      {\bf 62}, 024025 (2000);
                 [Erratum-ibid.\  D {\bf 62}, 109901 (2000)];
   S.~Frittelli, T.~P.~Kling and E.~T.~Newman,
             {\it ibid}   {\bf 61}, 064021 (2000).
 
\bibitem{MajuMukh} 
    A.~S.~Majumdar and N.~Mukherjee,
           Mod.\ Phys.\ Lett.\  A {\bf 20}, 2487 (2005);  
             Int.\ J.\ Mod.\ Phys.\  D {\bf 14}, 1095 (2005);  
    N.~Mukherjee and A.~S.~Majumdar, 
                    Gen.\ Rel.\ Grav.\  {\bf 39}, 583 (2007).
   
\bibitem{KP05}    
     C.~R.~Keeton and A.~O.~Petters, 
                  Phys.\ Rev.\  D {\bf 72}, 104006 (2005).
\bibitem{KPIW}  
      C.~R.~Keeton and A.~O.~Petters, 
                      Phys.\ Rev.\  D , {\bf 73}, 044024 (2006);
                      {\it ibid}, {\bf 73}, 104032 (2006);
    A.~O.~Petters,
                Mon.\ Not.\ Roy.\ Astron.\ Soc.\  {\bf 338}, 457 (2003); 
    S.~V.~Iyer and A.~O.~Petters,
                Gen.\ Rel.\ Grav.\  {\bf 39}, 1563 (2007);
    M.~C.~Werner and A.~O.~Petters,  Phys.\ Rev.\  D {\bf 76}, 064024 (2007).
       
\bibitem{ExoticLens}
        A.~Bhadra, Phys.\ Rev.\  D {\bf 67}, 103009 (2003);
        R.~Whisker, {\it ibid} {\bf 71}, 064004 (2005);
        K.~K.~Nandi, {\it ibid} {\bf 74},024020 (2006);
        G.~N.~Gyulchev and S.~S.~Yazadjiev,
                       {\it ibid} {\bf 75}, 023006 (2007);
        M.~Safonova and D.~F.~Torres,
                  Mod.\ Phys.\ Lett.\  A {\bf 17}, 1685 (2002).
\bibitem{BHLens}
     S.~Fernando and S.~Roberts,  Gen.\ Rel.\ Grav.\  {\bf 34}, 1221 (2002);
     C.~Stornaiolo, Gen.\ Rel.\ Grav.\  {\bf 34}, 2089 (2002);
     K.~Lake, Phys.\ Rev.\  D {\bf 65}, 087301 (2002);
     S.~E.~Vazquez and E.~P.~Esteban, Nuovo Cim.\  {\bf 119B}, 489 (2004);
     F.~Finelli, M.~Galaverni and A.~Gruppuso, Phys.\ Rev.\  D {\bf 75}, 043003 (2007).

\bibitem{Scalar}
     M.~P.~Dabrowski and F.~E.~Schunck, Astrophys.\ J.\  {\bf 535}, 316 (2000);
     T.~Matos and R.~Becerril, Class.\ Quant.\ Grav.\  {\bf 18}, 2015 (2001);
     V.~A.~De Lorenci, N.~Figueiredo, H.~H.~Fliche and M.~Novello,
                             Astron.\ Astrophys.\  {\bf 369}, 690 (2001);
     F.~E.~Schunck, B.~Fuchs and E.~W.~Mielke,Mon.\ Not.\ Roy.\ Astron.\ Soc.\  {\bf 369}, 485 (2006);     
     K.~Sarkar and A.~Bhadra, Class.\ Quant.\ Grav.\  {\bf 23}, 6101 (2006).

\bibitem{Math} S. Wolfram, MATHEMATICA 6.0.

\bibitem{Vir97}
       K.~S.~Virbhadra, Int. J. Mod. Phys. A  {\bf 12}, 4831 (1997).

\bibitem{VJJ97}
      K.~S.~Virbhadra, S. Jhingan, and P.~S.~Joshi, Int. J. Mod. Phys. D
            {\bf 6}, 357 (1997).
                           
\bibitem{SFE92}  P.~Schneider, J.~Ehlers,  and E.~E.~Falco,
            {\em Gravitational Lenses} (Springer-Verlag, Berlin, 1992).
\bibitem{Wein72} S. Weinberg, {\em  Gravitation and Cosmology: Principles  and
               Applications  of the General  Theory of Relativity} (Wiley, New York, 1972).
\bibitem{Janis-Newman-WinicourWyman}
      A.~I.~Janis, E.~T.~Newman, and J.~Winicour,  Phys. Rev. Lett. {\bf 20}, 878 (1968);
      M.~Wyman, Phys. Rev. D {\bf 24}, 839 (1981);
      M.~D.~Roberts, Astrophysics \& Space Sc. {\bf 200}, 331 (1993).

\bibitem{Eisenetal05} F.~Eisenhauer {\it et al.}, Astrophys. J. {\bf 628}, 246 (2005).
\bibitem{CKN07} A.~B.~Congdon, C.~R.~Keeton, and C.~E.~Nordgren, Phys.\ Rev.\  D {\bf 76} (2007), to appear.
\end{references}
\end{document}